\documentclass[journal]{IEEEtran}
\usepackage[english]{babel}
\usepackage[utf8]{inputenc}

\usepackage{amsmath}
\usepackage{amssymb}
\usepackage{mathtools}
\usepackage[cmintegrals]{newtxmath}
\usepackage{bbm}

\usepackage{booktabs}
\usepackage{tablefootnote}
\usepackage{multirow}

\usepackage{url}

\usepackage{pgfplots}
\pgfplotsset{compat=newest}
\usetikzlibrary{plotmarks}
\usetikzlibrary{arrows,decorations.markings,decorations.text,shapes,positioning,chains}
\usepgfplotslibrary{patchplots}
\usepackage{grffile}
\pgfplotsset{plot coordinates/math parser=false}
\newlength\fheight
\newlength\fwidth

\usepackage[font=footnotesize]{caption}

\usepackage{dblfloatfix}    

\newcommand{\intd}{\partial}  
\newcommand{\prr}{{\rm Pr}} 
\newcommand{\tth}{{\rm th}}
\newcommand{\ttx}{{\rm tx}}
\newcommand{\rrx}{{\rm rx}}
\newcommand{\nttx}{{\rm ntx}}
\newcommand{\ap}{{\rm AP}}
\newcommand{\acc}{{\rm acc}}
\newcommand{\st}{{\rm S}}
\newcommand{\R}{{\mathcal{R}}}

\newcommand{\E}[1]{\mathrm{E}\left[#1 \right]}

\usepackage{color}

\usepackage[acronym]{glossaries}
\newacronym{mac}{MAC}{Medium Access Control}
\newacronym{mc}{MC}{Markov Chain}
\newacronym{mcs}{MCS}{Modulation and Coding Scheme}
\newacronym{dcf}{DCF}{Distributed Coordination Function}
\newacronym{edca}{EDCA}{Enhanced Distributed Channel Access}
\newacronym{qos}{QoS}{Quality of Service}
\newacronym{qo}{QO}{quasi-omnidirectional}
\newacronym{csma}{CSMA}{Carrier Sense Multiple Access}
\newacronym{cbap}{CBAP}{Contention Based Access Period}
\newacronym{sp}{SP}{Service Period}
\newacronym{sta}{STA}{station}
\newacronym{ap}{AP}{Access Point}
\newacronym{bhi}{BHI}{Beacon Header Interval}
\newacronym{bi}{BI}{Beacon Interval}
\newacronym{rts}{RTS}{Request-To-Send}
\newacronym{cts}{CTS}{Clear-To-Send}
\newacronym{csmaca}{\gls{csma}/CA}{\gls{csma} with Collision Avoidance}
\newacronym{snr}{SNR}{Signal-to-Noise-Ratio}
\newacronym{mimo}{MIMO}{Multiple-Input Multiple-Output}
\newacronym{sifs}{SIFS}{Short Interframe Space}
\newacronym{difs}{DIFS}{Distributed Interframe Space}
\newacronym{ber}{BER}{Bit Error Rate}
\newacronym{eirp}{EIRP}{Equivalent Isotropic Radiated Power}
\newacronym{fcc}{FCC}{Federal Communication Commission}
\newacronym{etsi}{ETSI}{European Telecommunications Standards Institute}
\newacronym{psd}{PSD}{Power Spectral Density}
\newacronym{ppp}{PPP}{Poisson Point Process}
\newacronym{vbr}{VBR}{Variable Bit Rate}
\newacronym{dti}{DTI}{Data Transmission Interval}
\newacronym{ehf}{EHF}{Extremely High Frequency}
\newacronym{pbss}{PBSS}{Personal Basic Service Set}
\newacronym{pcp}{PCP}{\gls{pbss} Control Point}
\newacronym{sls}{SLS}{Sector-Level Sweep}
\newacronym{brp}{BRP}{Beam Refinement Protocol}
\newacronym{bti}{BTI}{Beacon Transmission Interval}
\newacronym{abft}{A-BFT}{Association-Beamforming Training}
\newacronym{ati}{ATI}{Announcement Transmission Interval}
\newacronym[\glslongpluralkey={Transmission Opportunities}]{txop}{TXOP}{Transmission Opportunity}
\newacronym{nav}{NAV}{Network Allocation Vector}

\begin{document}

\title{An Analytical Model for CBAP Allocations \\in IEEE 802.11ad}

\author{\IEEEauthorblockN{
Chiara Pielli, Tanguy Ropitault, Nada Golmie, and Michele Zorzi}\\
\thanks{Chiara Pielli (piellich@dei.unipd.it) and Michele Zorzi (zorzi@dei.unipd.it) are with the Department of Information Engineering, University of Padova, Padova, Italy.}
\thanks{Tanguy Ropitault (tanguy.ropitault@nist.gov) and Nada Golmie (nada.golmie@nist.gov) are with the National Institute of Standards and Technology (NIST), Gaithersburg, MD, US.}
}

\IEEEtitleabstractindextext{%
\begin{abstract}
The IEEE 802.11ad standard extends WiFi operation to the millimeter wave frequencies, and introduces novel features concerning both the physical (PHY) and \gls{mac} layers. However, while there are extensive research efforts to develop mechanisms for establishing and maintaining directional links for mmWave communications, fewer works deal with transmission scheduling and the hybrid \gls{mac} introduced by the standard. The hybrid \gls{mac} layer provides for two different kinds of resource allocations: \glspl{cbap} and contention free \glspl{sp}. In this paper, we propose a Markov Chain model to represent \glspl{cbap}, which takes into account operation interruptions due to scheduled \glspl{sp} and the deafness and hidden node problems that directional communication exacerbates. We also propose a mathematical analysis to assess interference among stations. 
We derive analytical expressions to assess the impact of various transmission parameters and of the Data Transmission Interval configuration on some key performance metrics such as throughput, delay and packet dropping rate. This information may be used to efficiently design a transmission scheduler that allocates contention-based and contention-free periods based on the application requirements. 
\end{abstract}

}

\maketitle
\IEEEdisplaynontitleabstractindextext
\IEEEpeerreviewmaketitle


\glsresetall

\section{Introduction}

Ratified in December $2012$, the IEEE 802.11ad amendment to the IEEE 802.11 standard targets short range millimeter wave (mmWave) communications in local area networks~\cite{802.11ad}. It is the fist amendment of 802.11 that concerns mmWaves, and more specifically it targets the $60$ GHz ISM unlicensed band~\cite{802.11ad}. 
MmWaves have been recently gaining a lot of momentum in telecommunications thanks to the wide spectrum available, which allows for channels with higher capacity and has the potential to eliminate the congestion issues encountered in the overcrowded sub-$6$-GHz bands.

The propagation environment in the mmWave spectrum is significantly different from that at sub-$6$-GHz frequencies, and is characterized by a high propagation loss and a significant sensitivity to blockage.  Blockage refers to very high attenuation due to obstacles; the human body for example could yield a penetration loss in the $20-35$ dB range~\cite{rangan2014millimeter}. 
Note, however, that the high attenuation may be an advantage for applications with short range, since it makes interference from farther transmissions negligible. 
The coverage range can be increased through beamforming, by focusing the power (both in transmission and in reception) towards the chosen direction, yielding a so-called directional link. This can be obtained by properly steering the elements of the antenna arrays. Also, the antenna arrays can be extremely compact and easily embedded into sensors and handsets, since the inter-antenna distance is proportional to the signal wavelength. 
Directional communication significantly attenuates interference among concurrent transmissions, yielding high potential for spatial sharing. Beamforming is a delicate process, and requires efficient beamforming training and beam tracking algorithms to establish and also maintain directional links, since poorly trained beams lead to extreme throughput loss: using the lowest \gls{mcs} defined by IEEE 802.11ad yields a drop of almost $95\%$ compared to the highest achievable data rate of $6.76$ Gbps~\cite{nitsche2014ieee}.

Because of the characteristics of the mmWave propagation environment, protocols designed for lower frequencies cannot simply be transposed to the mmWave band, but major design changes are required for both PHY and \gls{mac} layers.
While extensive research is ongoing to develop efficient beamforming training and beam tracking mechanisms~\cite{kutty2016beamforming,satchidanandan2018trackmac}, it is also necessary to understand how to access the wireless medium and use the beamformed links efficiently to transmit data.
The \gls{mac} layer of 802.11ad presents several features which yield an outstanding scheduling flexibility: it is possible to have both contention-based and contention-free allocations, and an additional mechanism built on top of the defined schedule allows to dynamically allocate channel time in quasi real-time. 
However, the standard~\cite{802.11ad} only provides rules for channel access; to the best of our knowledge, efficient scheduling schemes that exploit this hybrid \gls{mac} layer and match each traffic pattern to the most appropriate allocation is yet to be developed.
To realize an adaptive scheduler able to optimally allocate the channel time resources and accommodate disparate \gls{qos} requirements, it is first necessary to assess the performance that can be obtained with the mechanisms available in 802.11ad.


A mathematical model allows to understand the tradeoffs between the various system parameters and how they affect the network performance.
However, building a complete model is extremely challenging because there are several components to consider. 
In this paper we only focus on the performance that can be obtained in \gls{cbap} allocations, taking into account the presence of \glspl{sp} allocations. This is intended to represent a first step in the process of understanding and characterizing the various types of allocations that can be used in 802.11ad with the ultimate goal of designing an efficient allocation scheduler able to cope with heterogeneous traffic patterns and requirements.
In particular, we propose a variation of Bianchi's seminal model for the \gls{dcf} mechanism in legacy WiFi networks~\cite{bianchi2000performance}. Such variation addresses the main novel features of the 802.11ad standard and, unlike most of the works proposed in the literature, takes into account the deafness and hidden node problems, which are exacerbated by directional transmissions~\cite{akhtar2018directional}.  
Our model is based on a division of the area around a considered \gls{sta} into regions, similarly to what done in~\cite{chen2013directional}: \glspl{sta} belong to different groups based on whether they can overhear the messages sent by the \gls{sta} to and/or received from the \gls{ap}, according to their respective positions and beams. 
However,~\cite{chen2013directional} does not specify how to determine such regions; we instead explain how to compute the area of the regions mathematically, providing also the formulation for its expectation over the location of the considered station.
This classification of STAs is needed to characterize the probability of collision, and thus to evaluate performance metrics such as throughput, latency and dropping rate. Note that, although in directional communication systems the STAs rarely interfere with each other, the carrier sensing mechanism is not as effective, causing STAs to access the channel while the \gls{ap} is already involved in ongoing communications with other STAs.

The rest of the paper is structured as follows. Sec.~\ref{sec:protocol} gives an overview of the 802.11ad standard. Sec.~\ref{sec:scheduling} explains the scheduling problem and introduces the related works. The proposed model and the metrics used to evaluate the performance are described in Secs.~\ref{sec:model} and~\ref{sec:metric}, respectively. Sec.~\ref{sec:directional_model} explains how to compute the interference regions when constant-gain beam shapes are used. Sec.~\ref{sec:results} shows the numerical evaluations and, finally, Sec.~\ref{sec:conclusions} concludes the paper.

\section{802.11ad} \label{sec:protocol}

In this section we briefly describe the 802.11ad standard, with a special focus on the data transmission mechanisms.

\begin{figure}[t]
  \centering
  \includegraphics[width=\columnwidth]{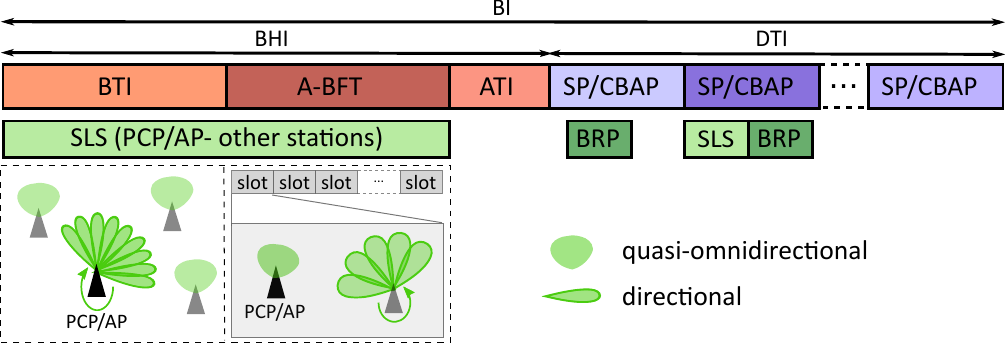}
  \caption{Structure of a BI. Green boxes correspond to beamforming training operations. The BHI is used for SLS with the PCP/AP: during the BTI, the PCP/AP trains its transmitting antenna pattern; during the A-BFT the other stations train their transmitting or receiving antenna patterns in dedicated slots. During the DTI, stations can perform both SLS and BRP phases with the PCP/AP and with other stations~\cite{pielli2018potential}.}
    \label{fig:BI}
\end{figure}

\subsection{Physical layer} 

The nominal channel bandwidth in 802.11ad is $2.16$ GHz, and there are up to $4$ channels in the ISM band around $60$ GHz, although channel availability varies from region to region. Only one channel at a time can be used for communication.
There are $32$ different \glspl{mcs} available, grouped into three different PHY layers, namely Control PHY, Single Carrier PHY and OFDM PHY, which differ for robustness, complexity, and achievable data rates. The standard also includes an energy-saving mode for battery-operated devices, which uses the low power Single Carrier PHY.

\subsection{Beamforming training} 
 
802.11ad introduces the concept of antenna sectors, which correspond to a discretization of the antenna space and reduce the number of possible beam directions to try. The standard supports up to four transmitter antenna arrays, four receiver antenna arrays, and 128 sectors per device. 
Beamforming training is realized in two subsequent stages: the \gls{sls} phase and the \gls{brp} phase, that aim at setting up a link between the stations and maximizing the gain, respectively. This mechanism can be performed between two STAs or a STA and the PCP/AP;~\footnote{Besides the traditional WiFi network topology, 802.11ad can also be used for \glspl{pbss}, i.e., network architectures for ad hoc modes. The central coordinator of 802.11ad networks can then be either a \gls{pcp} or an \gls{ap}; accordingly, it is generally denoted as PCP/AP to include both infrastructures.} the two devices are denoted as initiator and responder depending on who starts the SLS phase. 
The initiator sequentially tries different transmit antenna sector configurations, while the responder has its received antennas configured in a quasi-omnidirectional pattern and gives a feedback on each sector tried by the initiator, so as to determine the best coarse-grained antenna sector configuration. The SLS phase can be used also to inspect different configuration of the receiving antenna sectors: in this case, the initiator transmits multiple frames on the best known sector and the pairing node switches receiving sector. 
After a directional link has been established, the BRP phase is used to inspect narrower beams and, possibly, to optimize the antenna weight vectors in case of phased antenna arrays.
Since the BRP phase follows the SLS one, a reliable frame exchange is ensured.


\subsection{Beacon Intervals} 
Medium access time is divided into \glspl{bi}, with maximum duration of $1$ s (but typically chosen around $100$ ms~\cite{nitsche2014ieee}). Each BI consists of a \gls{bhi} and a \gls{dti}, as shown in Fig.~\ref{fig:BI}. 
The BHI replaces the single beacon frame of legacy WiFi networks and is used for synchronization and network management operations and to establish and maintain directional communication links between the STAs and the PCP/AP through beamforming training and beam tracking mechanisms; the DTI is used for data transmission and for beamforming training with the PCP/AP and between STAs. 

The \gls{bhi} includes up to three access periods, all of them optional: the \gls{bti}, the \gls{abft}, and the \gls{ati}.
The \gls{bti} is used for SLS beamforming training of the PCP/AP's antennas and network announcement: the PCP/AP broadcasts beacon frames iterating through different sectors, performing the first part of the SLS phase with the STAs, which have their receiving antennas configured in a quasi-omnidirectional pattern since they do not know a-priori the direction to use to receive the beacons. 
The SLS phase started in the BTI is completed in the A-BFT, which is divided into slots during which STAs separately train their antenna sectors for communication with the PCP/AP, and provide feedback to the PCP/AP about the sector to use for transmitting to them. 
Finally, the \gls{ati} is used to exchange management information between the PCP/AP and associated and beamtrained stations, such as resource requests and allocation information for the DTI. 


\subsection{Data transmission} \label{sec:dti}
The DTI is made up of contention-free \glspl{sp} for exclusive communication between a dedicated pair of nodes\footnote{Technically, spatial sharing allows communication for multiple pair of nodes, but interference among pairs is checked to be basically null.} and \glspl{cbap} where stations compete for access. SPs and CBAPs can be in any number and combination, and their scheduling is advertised by the PCP/AP through beacons in the BTI and/or specific frames in the ATI. An allocation is defined by several fields, including the type of allocation (SP or CBAP), the addresses of the source and destination \glspl{sta} involved in the allocation (which can be unicast, multicast or broadcast), its total duration and starting time and the number of blocks it is made of, beamforming training information if needed, and whether the allocation is pseudostatic, meaning that it recurs in subsequent \glspl{bi}~\cite{802.11ad}.
Note that this schedule is set up prior to the beginning of the \gls{dti}. 
In addition, a dynamic channel time allocation mechanism allows STAs to reserve channel time in almost real-time over both SPs and CBAPs. 


\textit{\textbf{Contention-based access.}} CBAPs follow the \gls{edca} mechanism, which is an enhanced \gls{dcf} that includes functionalities to handle traffic categories with different priorities, frame aggregation and block acknowledgments.
Stations compete for access and can obtain \glspl{txop} (contention-free periods) by winning an instance of \gls{edca} contention or by receiving a Grant frame; the \gls{txop} duration depends on the traffic category.

The \gls{dcf} is based on \gls{csmaca}: before transmission, the channel needs to be sensed idle for a minimum amount of time, namely a \gls{difs}. If the channel is sensed busy, the transmission is postponed: the \gls{sta} picks a backoff counter uniformly distributed in $\{0,\ldots,W_i-1\}$, where $W_i$ is the size of the contention window at the $i$-th retransmission attempt. The contention window starts at a minimum value and doubles at each collision, until it saturates to a maximum value. The backoff time counter is decremented as long as the channel is sensed idle, frozen when a transmission is detected on the channel or the \gls{cbap} operation is suspended, and reactivated when the channel is sensed idle again for at least a DIFS (after that the CBAP operation has been resumed). When the backoff counter expires, the \gls{sta} accesses the channel.

In 802.11ad, the channel status is determined through a combined physical and virtual carrier sensing; the former consists in energy or preamble detection over the channel, the latter is realized through \glspl{nav}. The \glspl{nav} are counters based on the transmission duration information announced in \gls{rts} and \gls{cts} frames prior to the actual exchange of data and maintain a prediction of future traffic on the medium. 

The directional nature of communication at mmWaves makes the carrier sensing operations problematic~\cite{nitsche2014ieee} because there may be possible interference even though the medium was considered to be idle.

\textit{\textbf{Contention-free access.}}
\glspl{sp} are contention-free periods assigned by the PCP/AP for exclusive communication between a pair of \glspl{sta}.
The directional communication enables the possibility of spatial sharing, i.e., simultaneous SPs involving different STAs can be scheduled, provided that they do not interfere with each other; this requires a preliminary interference assessment phase, which is coordinated by the PCP/AP. Note that building and updating the interference map may result in huge overhead in case of mobility.


\textit{\textbf{Dynamic allocation mechanism.}}
This mechanism is built on scheduled SPs and CBAPs with specific configuration and enables near-real-time reservation of channel time; the dynamic allocations do not persist beyond a BI.
Stations can be polled by the PCP/AP and ask for channel time, which will be granted back to back. 

The dynamic mechanism also includes the possibility of truncating and extending SPs, to exploit unused channel time and finalize the ongoing communication without additional delay and scheduling, respectively. When an SP is truncated,  either the relinquished channel time is used as a CBAP or the PCP/AP polls \glspl{sta} so that they can ask for channel time.

\textit{\textbf{Resource scheduling.}}
Evidently, there are many elements that need to be taken into account to appropriately schedule the \gls{dti} based on the \gls{qos} requirements. 
In addition to modeling data transmission in both \gls{cbap} and \gls{sp} allocations, it is necessary to understand in which cases the dynamic allocation mechanism yields better performance than the predefined schedule. Another aspect that should be taken into consideration is power consumption: the presence of energy constrained devices may require changes in the scheduling, e.g., in the allocation order or by assigning more \gls{sp} allocations. A critical issue is represented by the beamforming training (see Sec.~\ref{sec:beams}) which introduces overhead and may degrade the network performance.
Mainly, there are three knobs available to the protocol designer: 

\begin{itemize}
 \item \emph{Contention-based or contention free allocation.} This is the most meaningful choice as it impacts the way the medium is accessed and thus plays a direct role on the performance. \gls{sp} allocations grant dedicated resources and the obtained performance only depends on the channel status, being therefore more predictable than when interference comes into play. Clearly, setting up the scheduled sessions introduces overhead and some latency, but the beam steering process is simplified since the receiver knows who is going to transmit and can steer its receiving beam towards the sender, and the \glspl{sta} not involved in the \gls{sp} can go to sleep and save power. On the other hand, \glspl{cbap} are distributed and robust and good for unpredictable bursty traffic. Nonetheless, carrier sensing may be problematic due to the use of directional antennas. 
 Also, during CBAP period, STAs cannot go into power saving mode due to the inner nature of CSMA/CA.
 SP allocations are particularly suitable for periodic reporting with \gls{qos} demands, but CBAPs may be preferable in case of less stringent \gls{qos} requirements because channel resources are available to more stations.
 
 \item \emph{Pseudo static allocation.} In this way, it is possible to decide whether the allocation will recur in successive \glspl{bi}. This is very useful for predictable traffic patterns as it avoids the need to schedule the allocation every time and limits the signaling overhead. 
 
 \item \emph{Dynamic allocation.} It allows quasi-real-time channel use, but has a polling overhead and the scheduled allocation over which it is applied needs to satisfy certain conditions. This feature can be useful for unpredictable transmissions that need to be delivered with specific \gls{qos} requirements. 
\end{itemize}

\section{Related work}

The seminal work of~\cite{bianchi2000performance} introduces a \gls{mc} model of the IEEE 802.11 \gls{dcf}. Although several variations on such model have been proposed to account for, e.g., finite number of retransmissions~\cite{chatzimisios2003ieee}, heterogeneous \gls{qos}~\cite{robinson2004saturation} and hidden node problem~\cite{hung2010performance}, none of them can be readily applied to the hybrid \gls{mac} layer of IEEE 802.11ad, as different changes are needed to account for its peculiar features.

Some works in the literature propose adaptations of Bianchi's model for 802.11ad. Most of them, however, do not model the effect of directional communication properly, as they neglect the deafness and hidden node problems.
For example, \cite{chandra2017performance} uses a $3$-dimensional \gls{mc} model to analyze the channel utilization and the average \gls{mac} layer delay that can be obtained in \glspl{cbap}. This model accounts for the presence of allocations other than \gls{cbap} and for the fact that backoff counters are frozen when \gls{csmaca} operation is suspended. However, it does not introduce the maximum contention window size, so that the contention window keeps doubling at each retransmission stage. Moreover, the model assumes that \glspl{cbap} are allocated to sectors, so that two \glspl{sta} belonging to different sectors cannot compete for the channel time in the same allocation. According to the standard~\cite{802.11ad}, this is not necessarily true, since any subset of stations can participate in a \gls{cbap}, with potential deafness and hidden node issues. The model also erroneously assumes that all \glspl{sta} in the same sector can overhear the messages that other nodes exchange with the \gls{ap}. Thus, the assumption made in~\cite{chandra2017performance} strongly affects the analysis of the delay and the impact of the number of sectors used by the PCP/AP on the system performance, as the role of directional transmissions and deafness is neglected.

Similar assumptions have been made in~\cite{hemanth2013performance}, which models \glspl{cbap} with a $2$-dimensional \gls{mc} for unsaturated sources considering also the contention-free allocations of 802.11ad. However, besides neglecting the deafness problem, the model assumes that the \gls{dti} is made of \gls{sp} allocations followed by a single \gls{cbap} allocation at the end of the \gls{dti}, while the standard~\cite{802.11ad} envisages \gls{sp} and \gls{cbap} allocations in any number and order. This assumption may strongly affect the delay, as different configurations of the \gls{dti} may yield different performance.
Also~\cite{rajan2016saturation} uses a  $2$-dimensional \gls{mc} to analyze the saturation throughput in \gls{cbap} but neglects the deafness issues and assumes the same specific configuration of the \gls{dti} as in~\cite{hemanth2013performance}.

A more accurate approach to directional communication in WiFi networks is presented in~\cite{babich2009throughput}, which however is not designed for 802.11ad so that it does not consider the presence of \gls{sp} allocations and the related backoff counter freezing. The model considers an accurate model for directional transmission, with the presence of side lobes with small antenna gain and corresponding regions with different levels of interference.
Also~\cite{chen2013directional} takes into account deafness and hidden node problems, and subdivides the area around a \gls{sta} based on the interference level; \glspl{cbap} are then modeled using a $3$-dimensional \gls{mc}.

Other works in the literature consider different aspects of the \gls{dti}. For example, \cite{rajan2017theoretical} derives the theoretical maximum throughput for \glspl{cbap} when two-level \gls{mac} frame aggregation is used. \cite{akhtar2018directional} proposed a directional MAC protocol to be used on top of 802.11ad: it allows the use of sequential directional \gls{rts} messages that a STA sends in all directions and that can therefore be overheard by all other \glspl{sta}. The beamforming issue is considered in~\cite{shokri2015beam}, which proposes a joint optimization of beamwidth selection and scheduling to maximize the effective network throughput.

For what concerns \glspl{sp}, an accurate mathematical model for their preliminary allocation is presented in \cite{khorov2016mathematical}. It considers the presence of quasi-periodic structures with multiple blocks within the same allocation, the erroneous nature of the wireless medium, and the possibility of multiple consecutive transmissions within the same allocation. A $3$-dimensional MC is used to model a \gls{vbr} flow with packets arriving in batches of random size at regular intervals and can be used to derive the optimal SP allocation that satisfies the \gls{qos} requirement.

\section{System model} \label{sec:model}

We now introduce our analytical model for CBAP operation in 802.11ad.
We denote as $T_{\rm BI}$ the duration of a \gls{bi} and as $T_{\rm BHI}, T_{\rm CBAP}$ and $T_{\rm SP}$ the time dedicated to \gls{bhi}, \glspl{cbap} and \glspl{sp} during a \gls{bi}, respectively. The total time $T_{\rm CBAP}$ dedicated for contention-based access in a \gls{bi} is distributed among $N_{\rm CBAP}$ allocations with the same duration, while $T_{\rm SP}$ is distributed among $N_{\rm SP}$ allocations with the same duration. 

We make two assumptions: i) all STAs in the network implement a single Access Category (AC) only, hence service differentiation is not considered, and ii) the beamforming training has already been performed, so that the \glspl{sta} already know how to steer their antennas to communicate with the \gls{ap}.
Also, we only focus on the classic WiFi network where a certain number of \glspl{sta} communicate solely with the \gls{ap}; we consider that the \gls{rts}/\gls{cts} mechanism is used.

To assess the performance that can be obtained in a \gls{cbap}, we leverage on Bianchi's seminal work~\cite{bianchi2000performance} and adapt it to model the features of \glspl{cbap} in 802.11ad.
First of all we explain how directionality affects the communication during the contention-based channel access, then we describe the proposed model, and finally we discuss the performance metrics used in the numerical evaluation.

\subsection{Directional communication in CBAPs} \label{sec:dir_comm}

Besides the need of beamforming training and beam tracking mechanisms, the directional nature of communication in 802.11ad implies substantial changes also from a data transmission perspective. As explained in Sec~\ref{sec:dti}, \glspl{cbap} are based on the \gls{edca}; however, the traditional approaches used in the literature  need to be adapted to take directionality into account, since the consequent deafness and hidden node problems may significantly affect the system performance.

The most widely used approach in the literature to model the \gls{dcf} and \gls{edca} mechanisms is Bianchi's model~\cite{bianchi2000performance}. It takes the perspective of a target node and models the backoff process as a two dimensional \gls{mc}, where state $(i,k)$ refers to the $i^\tth$ backoff stage with the backoff counter $k \in \{0,\ldots,W_i-1\}$, where $W_i$ is the duration of the contention window at the $i^\tth$ retransmission attempt. The counter is decremented with probability $1$ whenever the channel is sensed idle; when it reaches $0$, the \gls{sta} attempts to transmit. The time spent in each state depends on what happens in the channel  meanwhile, as it may be idle, used for a successful transmission, or used simultaneously by colliding \glspl{sta}.
The original model was proposed for omnidirectional communication, so that each \gls{sta} is aware of ongoing transmissions and can defer its own when it senses the channel to be idle. Collisions only happen when multiple \glspl{sta} access the channel simultaneously because their backoff counters expired (at least two \glspl{sta} are in a state $(\cdot,0)$).

\begin{figure}[t]
  \centering
  \includegraphics[width=\columnwidth]{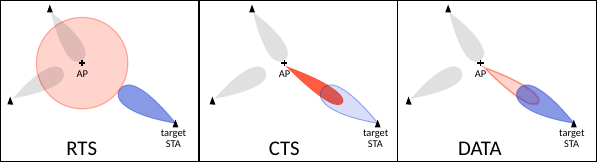}
  \caption{Communication phases between the AP and a target STA and two other STAs that listen in direction of the AP. Darker beams indicate a transmission, while lighter ones indicate that the device is listening. The target STA directionally transmits the RTS while the AP is listening in QO mode; then the AP steers its transmitting antennas towards the target STA and sends the CTS to the target STA, which then replies transmitting a data message.}
    \label{fig:phases}
\end{figure}

In the case of directional communication, however, \glspl{sta} may not hear ongoing transmissions, resulting in a much higher collision probability. 
In this work, we assume that the \gls{rts}/\gls{cts} mechanism is used. 
Since a \gls{sta} communicates only with the \gls{ap}, it always has both its transmitting and receiving antenna patterns configured towards it. 
The \gls{ap} instead listens to the channel in a \gls{qo} mode, and, upon the reception of an RTS, it switches its antenna configuration to point towards the \gls{sta} that sent it. Fig.~\ref{fig:phases} illustrates the direction of the various phases of the communication between a \gls{sta} and the \gls{ap}. Note that the messages can be heard only by a limited number of other \glspl{sta}. The received power $P_\rrx$ at a \gls{sta} is in fact
\begin{equation} \label{eq:p_rx}
 P_\rrx = P_\ttx \frac{g_\ttx(\theta_\ttx, \varphi_\rrx) g_\rrx(\theta_\ttx, \varphi_\rrx)}{A d^\eta}
\end{equation}
where $P_\ttx$ is the power used to transmit, $d$ is the distance from the transmitter, $\eta$ is the path-loss exponent, $A$ is a normalizing path-loss term, and $g_\ttx$ and $g_\rrx$ are the antenna gains of the transmitter and receiver, respectively. They both depend on the direction of the antennas with respect to the line of sight between the two \glspl{sta}, thus on the angles $\theta_\ttx$ and $\varphi_\rrx$. If the gains are very small, $P_\rrx$ may be too low in order for the receiver to decode the signal properly.

Consider a network consisting of $n$ \glspl{sta} and a target \gls{sta} that communicates with the \gls{ap}, so that the \gls{sta} and the \gls{ap} point to each other, and the antenna gains in the other directions are minimal. It is possible to cluster the other $n-1$ \glspl{sta} into four groups:
\begin{itemize}
 \item $n_{I,1}$: \glspl{sta} that can overhear the messages sent from the target \gls{sta} to the \gls{ap} but not those sent from the \gls{ap} to the \gls{sta}.
 
 \item $n_{I,2}$: \glspl{sta} that can overhear the messages from the \gls{ap} to the target \gls{sta} but not those from the \gls{sta} to the \gls{ap}.
 
 \item $n_{I,3}$: \glspl{sta} that can overhear the whole communication  between the \gls{ap} and the target \gls{sta}.
 
 \item $n_{I,4}$: \glspl{sta} that cannot overhear any messages exchanged   between the \gls{ap} and the target \gls{sta}.
\end{itemize}

Analogously, from the perspective of a \gls{sta} that listens to the channel, the other \glspl{sta} can be divided into four groups $n_{O,1}$ (\glspl{sta} of which it can hear the messages to the \gls{ap} but not the messages that the \gls{ap} sends to them), $n_{O,2}$ (\glspl{sta} of which it cannot hear the messages to the \gls{ap} but can hear those that the \gls{ap} sends to them), $n_{O,3}$ (\glspl{sta} of which it can hear all the messages exchanged with the \gls{ap}), and $n_{O,4}$ (\glspl{sta} whose messages exchanged with the AP cannot be heard).

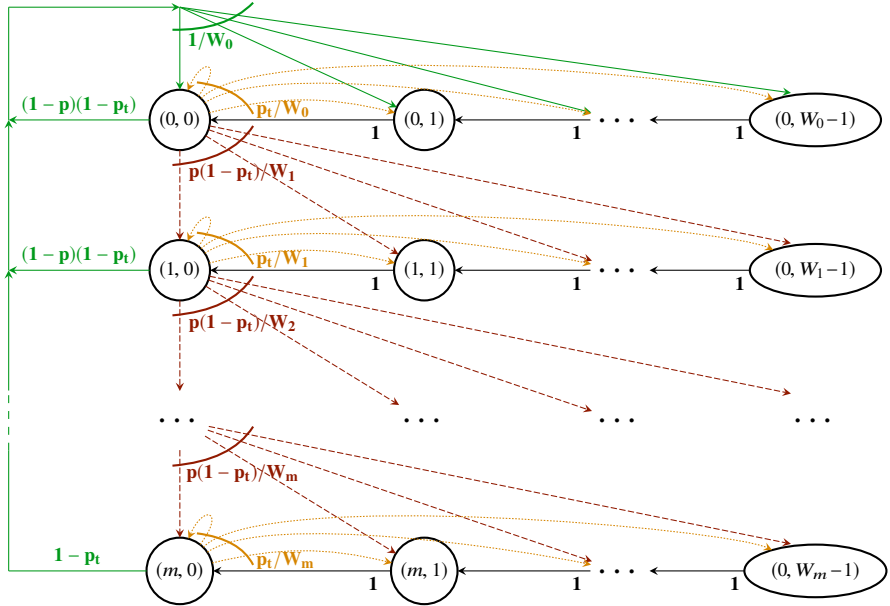
\begin{figure*}
	\captionsetup{justification=centering}
	\centering
	\vspace{-1.5em}\definecolor{darkgreen}{rgb}{0,0.6,0.1}%
\definecolor{darkred}{rgb}{0.57,0.1,0}%
\definecolor{darkblue}{rgb}{0,0.29,0.73}%
\definecolor{darkorange}{rgb}{0.84,0.52}%

\begin{tikzpicture}[xscale=1.3,yscale=0.8,outer sep=0cm]
\scriptsize
\tikzstyle{dotted} = [dash pattern=on \pgflinewidth off 0.5mm] 
\tikzstyle{dashed2} = [dash pattern=on 7.5*0.4*0.8pt off 7.5*0.15*0.8pt]
\tikzstyle{dashdotted} = [dash pattern=on 7.5*0.8*0.6pt off 7.5*0.8*0.3pt on \the\pgflinewidth off 7.5*0.8*0.3pt]
\tikzstyle{dotted2} = [dash pattern=on 7.5*0.1*0.8pt off 7.5*0.1*0.8pt]

\tikzset{>=stealth} 

\def \radius {.7cm}
\def \xspace {2.5cm}
\def \yspace {2.5cm}
\def \poshor {0.1}
\def \poscol {0.2}
\def \postimeout {0.2}

\node (0) at (1*\xspace,3.75*\yspace) {};

\node[draw, circle, minimum size=\radius,thick](00) at (1*\xspace,3*\yspace) {$(0,0)$};
\node[draw, circle, minimum size=\radius,thick](01) at (2*\xspace,3*\yspace) {$(0,1)$};
\node[draw=none, circle, minimum size=\radius,thick](02) at (2.8*\xspace,3*\yspace) {\Large\ldots};
\node[draw, ellipse, thick, minimum height=\radius, minimum width=0.1*\radius](03) at (3.6*\xspace,3*\yspace) {$(0,W_0\!-\!1)$};

\node[draw, circle, minimum size=\radius,thick](10) at (1*\xspace,2*\yspace) {$(1,0)$};
\node[draw, circle, minimum size=\radius,thick](11) at (2*\xspace,2*\yspace) {$(1,1)$};
\node[draw=none, circle, minimum size=\radius,thick](12) at (2.8*\xspace,2*\yspace) {\Large\ldots};
\node[draw, ellipse, thick, minimum height=\radius, minimum width=0.1*\radius](13) at (3.6*\xspace,2*\yspace) {$(0,W_1\!-\!1)$};

\node[draw=none, circle, minimum size=0.7*\radius](i0) at (1*\xspace,1*\yspace) {\Large\ldots};
\node[draw=none, circle, minimum size=\radius](i1) at (2*\xspace,1*\yspace) {\Large\ldots};
\node[draw=none, circle, minimum size=\radius](i2) at (2.8*\xspace,1*\yspace) {\Large\ldots};
\node[draw=none, circle, minimum size=\radius](i3) at (3.6*\xspace,1*\yspace) {\Large\ldots};

\node[use as bounding box](m) at (0.3*\xspace,0*\yspace) {};
\node[draw, circle, minimum size=\radius,thick](m0) at (1*\xspace,0*\yspace) {$(m,0)$};
\node[draw, circle, minimum size=\radius,thick](m1) at (2*\xspace,0*\yspace) {$(m,1)$};
\node[draw=none, circle, minimum size=\radius,thick](m2) at (2.8*\xspace,0*\yspace) {\Large\ldots};
\node[draw, ellipse, thick, minimum height=\radius, minimum width=0.1*\radius](m3) at (3.6*\xspace,0*\yspace) {$(0,W_m\!-\!1)$};

\draw[->] (03) to node[below,pos=\poshor] {$\mathbf{1}$} (02);
\draw[->] (02) to node[below,pos=\poshor] {$\mathbf{1}$} (01);
\draw[->] (01) to node[below,pos=\poshor] {$\mathbf{1}$} (00);
\draw[->] (13) to node[below,pos=\poshor] {$\mathbf{1}$} (12);
\draw[->] (12) to node[below,pos=\poshor] {$\mathbf{1}$} (11);
\draw[->] (11) to node[below,pos=\poshor] {$\mathbf{1}$} (10);
\draw[->] (m3) to node[below,pos=\poshor] {$\mathbf{1}$} (m2);
\draw[->] (m2) to node[below,pos=\poshor] {$\mathbf{1}$} (m1);
\draw[->] (m1) to node[below,pos=\poshor] {$\mathbf{1}$} (m0);

\draw[->,darkred,dashed2] (00) to (10);
\draw[->,darkred,dashed2] (00) to (11);
\draw[->,darkred,dashed2] (00) to (12);
\draw[->,darkred,dashed2] (00) to (3.5*\xspace,2.18*\yspace);
\draw[->,darkred,dashed2] (10) to (i0);
\draw[->,darkred,dashed2] (10) to (i1);
\draw[->,darkred,dashed2] (10) to (i2);
\draw[->,darkred,dashed2] (10) to (3.5*\xspace,1.18*\yspace);
\draw[->,darkred,dashed2] (i0) to (m0);
\draw[->,darkred,dashed2] (i0) to (m1);
\draw[->,darkred,dashed2] (i0) to (m2);
\draw[->,darkred,dashed2] (i0) to (3.5*\xspace,0.18*\yspace);
\draw[darkred,thick] (0.97*\xspace,2.7*\yspace) to[bend right]  node[below,pos=0.8,yshift=-0.25cm] {$\mathbf{p(1-p_t)/W_1}$}(1.3*\xspace,2.96*\yspace);
\draw[darkred,thick] (0.97*\xspace,1.7*\yspace) to[bend right]  node[below,pos=0.8,yshift=-0.25cm] {$\mathbf{p(1-p_t)/W_2}$}(1.3*\xspace,1.96*\yspace);
\draw[darkred,thick] (0.97*\xspace,0.7*\yspace) to[bend right]  node[below,pos=0.8,yshift=-0.25cm] {$\mathbf{p(1-p_t)/W_m}$}(1.3*\xspace,0.96*\yspace);

\draw[->,darkgreen] (00) -- node[midway,above] {$\mathbf{(1-p)(1-p_t)}$} (0.3*\xspace,3*\yspace);
\draw[->,darkgreen] (10) -- node[midway,above] {$\mathbf{(1-p)(1-p_t)}$} (0.3*\xspace,2*\yspace);
\draw[darkgreen] (m0) -- (0.3*\xspace,0*\yspace) node[midway, above] (edge) {$\mathbf{1-p_t}$} -- (0.3*\xspace,0*\yspace);
\draw[darkgreen] (0.3*\xspace,0*\yspace) -- (0.3*\xspace,0.8*\yspace);
\draw[dashed,darkgreen] (0.3*\xspace,0.8*\yspace)  -- (0.3*\xspace,1.2*\yspace);
\draw[->,darkgreen] (0.3*\xspace,1.2*\yspace)  -- (0.3*\xspace,2*\yspace);
\draw[->,darkgreen] (0.3*\xspace,2*\yspace)  -- (0.3*\xspace,3*\yspace);
\draw[darkgreen] (0.3*\xspace,3*\yspace)  -- (0.3*\xspace,3.75*\yspace);
\draw[->,darkgreen] (0.3*\xspace,3.75*\yspace) to (1*\xspace,3.75*\yspace);
\draw[->,darkgreen] (1*\xspace,3.75*\yspace) to (00);
\draw[->,darkgreen] (1*\xspace,3.75*\yspace) to (01);
\draw[->,darkgreen] (1*\xspace,3.75*\yspace) to (02);
\draw[->,darkgreen] (1*\xspace,3.75*\yspace) to (3.5*\xspace,3.18*\yspace);
\draw[darkgreen,thick] (.97*\xspace,3.6*\yspace) to[bend right]  node[below,pos=0.4] {$\mathbf{1/W_0}$}(1.3*\xspace,3.78*\yspace);

\draw[->,darkorange,dotted2] (00) to[distance=.6cm,in=75,out=60] (00);
\draw[->,darkorange,dotted2] (00) .. controls +(1,0.4) and +(-1,0.4) .. (01);
\draw[->,darkorange,dotted2] (00) .. controls +(.9,1) and +(-1,0.35) .. (02);
\draw[->,darkorange,dotted2] (00) .. controls +(0.85,1.4) and +(-0.9,0.7) .. (03);
\draw[darkorange,thick] (1.07*\xspace,3.25*\yspace) to[bend left]  node[right,pos=0.9] {$\mathbf{p_t/W_0}$}(1.3*\xspace,3.04*\yspace);
\draw[->,darkorange,dotted2] (10) to[distance=.6cm,in=75,out=60] (10);
\draw[->,darkorange,dotted2] (10) .. controls +(1,0.4) and +(-1,0.4) .. (11);
\draw[->,darkorange,dotted2] (10) .. controls +(.9,1) and +(-1,0.35) .. (12);
\draw[->,darkorange,dotted2] (10) .. controls +(0.85,1.4) and +(-0.9,0.7) .. (13);
\draw[darkorange,thick] (1.07*\xspace,2.25*\yspace) to[bend left]  node[right,pos=0.9] {$\mathbf{p_t/W_1}$}(1.3*\xspace,2.04*\yspace);
\draw[->,darkorange,dotted2] (m0) to[distance=.6cm,in=75,out=60] (m0);
\draw[->,darkorange,dotted2] (m0) .. controls +(1,0.4) and +(-1,0.4) .. (m1);
\draw[->,darkorange,dotted2] (m0) .. controls +(.9,1) and +(-1,0.35) .. (m2);
\draw[->,darkorange,dotted2] (m0) .. controls +(0.85,1.4) and +(-0.9,0.7) .. (m3);
\draw[darkorange,thick] (1.07*\xspace,0.25*\yspace) to[bend left]  node[right,pos=0.9] {$\mathbf{p_t/W_m}$}(1.3*\xspace,0.04*\yspace);


\end{tikzpicture}\vspace*{-1.5em}
	\caption{Macro Markov chain (adaptation of Bianchi's model~\cite{bianchi2000performance}).}
	\label{fig:macro_mc}
\end{figure*} 

Consequently, collisions can happen at three different stages of the uplink communication from a target \gls{sta} to the \gls{ap}.
\begin{enumerate}
 \item \label{case:1} The target \gls{sta} accesses the channel to transmit its \gls{rts}, but collides for sure. This can happen for three different reasons: i) if any other \gls{sta} accesses the channel at the same time, as in the legacy WiFi, ii) if a \gls{sta} belonging to groups $n_{O,2}$ or $n_{O,4}$ is transmitting the \gls{rts} to the \gls{ap}, or iii) if a \gls{sta} in group $n_{O,4}$ has already sent the RTS and is going on with the communication with the \gls{ap}. 
 Notice that, in the latter case, the transmission of the target \gls{sta} fails, because the \gls{ap} is listening in the direction of the \gls{sta} from group $n_{O,4}$\footnote{Different considerations can be made when considering \gls{mimo} systems, but this is out of the scope of this paper.}, but the ongoing data transmission may still be successful, as the directionality highly attenuates the interference. In this work we assume that, except for errors in the channel, the ongoing data transmission is successful.
 
 \item \label{case:2} If none of the previous conditions happened, the transmission of the \gls{rts} may still be vulnerable to interference. This happens when a \gls{sta} in groups $n_{I,2}$ or $n_{I,4}$ accesses the channel meanwhile. The packets will then collide.
 
 \item \label{case:3} If the transmission of the \gls{rts} was successful, the \gls{ap} sends the \gls{cts} and the target \gls{sta} can proceed with the data transmission. However, a \gls{sta} in group $n_{I,4}$ is unaware of the ongoing communication and may try to access the channel. As assumed in case  \ref{case:1}iii), the outcome of the ongoing transmission only depends on channel errors, while the \gls{sta} that accesses the channel will register a collision.
\end{enumerate}

In the remainder of this section, we propose an adaptation of Bianchi's model that accounts for the directionality of transmissions, assuming that the regions corresponding to the four groups of nodes are known; Sec.~\ref{sec:directional_model} introduces an analytical model to compute such regions when constant-gain beam shapes are used.

\setcounter{equation}{3}
\begin{table*}[t]
\begin{normalsize}
\begin{align} \label{eq:b00}
 b_{0,0} &= \begin{dcases}             
	  \frac{2(1-2p)(1-p)}{W_0(1-(2p)^{m+1}(1-p)+ (1-p^{m+1})(1-2p)} \qquad \qquad \qquad \qquad \qquad \qquad \qquad \;\;\: \text{if } m \le m'\\
	  \\
	  \frac{2(1-2p)(1-p)}{W_0(1-(2p)^{m'+1})(1-p) + 2^{m'} W_0 (p^{m'}-p^m)(1-2p)p + (1-p^{m+1})(1-2p)} \qquad \text{if } m > m'
            \end{dcases} 
\end{align}
\end{normalsize}
\rule{\textwidth}{0.4pt}
\end{table*}
\setcounter{equation}{1}

\subsection{Rethinking Bianchi's model}

Bianchi's model~\cite{bianchi2000performance} needs three major adaptations in order to be suitable for 802.11ad, which are caused by the following features.

\begin{enumerate}
 \item \label{case:mod_1} \glspl{cbap} can be interrupted because there is a scheduled \gls{sp} or the \gls{bhi} of the next \gls{bi}. In this case, all backoff counters have to freeze~\cite{802.11ad}; they will be restored in the next \gls{cbap}. This affects the time that a \gls{sta} spends in a state $(i,k), \,k\in\{1,\ldots,W_i-1\}$ before decrementing its backoff counter and transitioning to $(i,k-1)$; the transition probability from $(i,k)$ to $(i,k-1)$ is $1$ as in Bianchi's original model. We denote the freezing probability as $p_f$.
 
 \item \label{case:mod_2} The finite duration of \glspl{cbap} may also cause transmissions deferral. In fact, if the backoff counter of a \gls{sta} reaches $0$ but there is not enough time to complete a transmission, that \gls{sta} should refrain from transmitting. The standard \cite{802.11ad} however does not specify how to handle the backoff counter in this case. A possibile strategy consists in freezing it when it expires and then restoring it in the next \gls{cbap}, as in case \ref{case:mod_1}, so that the involved STA can transmit as soon as the \gls{edca} operation starts again. However, such approach may easily lead to collisions: if the counters of multiple \glspl{sta} expire during the time needed for a complete transmission at the end of the current \gls{cbap}, and they all get frozen to $0$, all such \glspl{sta} will attempt accessing the channel simultaneously in the next \gls{cbap}. To avoid this, we propose to use the same approach used in~\cite{hemanth2013performance}, so that a new backoff counter is randomly chosen from the current window (no collision happened). This causes the addition of new transitions from state $(i,0)$ to $(i,k),\,k\in\{0,\ldots,W_i-1\}$. 
 We denote the probability of insufficient time in the current \gls{cbap} as timeout probability $p_t$.
 
 \item \label{case:mod_3} As discussed in Sec.~\ref{sec:dir_comm}, the directional nature of mmWave communication has a huge impact on the operation of the \gls{dcf} mode because of deafness and the increased hidden node problem. This modifies the collision probability and the time spent in each state, which depend on the behavior of \glspl{sta} whose transmissions can be detected by the target \gls{sta}.
\end{enumerate}

Fig.~\ref{fig:macro_mc} represents the \gls{mc} we propose to model the behavior of a \gls{sta} during \gls{cbap} operation.
As in Bianchi's model, a state $(i,k),\,i\in\{0,\ldots m\},\,k\in\{0,\ldots,W_i-1\}$ refers to the $i^\tth$ backoff stage with the backoff counter being equal to $k$. Here, $m$ is the maximum number of retransmissions. The contention window in stage $i$ is $W_i=\min\{2^i W_0, 2^{m'} W_0\}$, where the initial window $W_0$ and the maximum window $2^{m'} W_0$ are defined in the standard.

From a state $(i,k),\,k\!>\!0$, the backoff counter is decremented with probability $1$ (solid black transitions in Fig.~\ref{fig:macro_mc}), but the time needed to transition to the next state $(i,k\!-\!1)$ is variable, depending on how the channel is being used. When it reaches a state $(i,0)$, the \gls{sta} might be constrained to defer its transmission (adaptation~\ref{case:mod_2}). The residual time in the current \gls{cbap} is uniformly distributed in $[0, T_{\rm CBAP}/N_{\rm CBAP}]$, where $T_{\rm CBAP}/N_{\rm CBAP}$ is the average duration of a \gls{cbap} allocation in the \gls{bi}. Then, the probability that there is no sufficient time to complete a transmission of duration $T_L$ can be approximated as
\begin{equation} \label{eq:p_t}
 p_t = \frac{T_L}{T_{\rm CBAP}/N_{\rm CBAP}}\,.
\end{equation}

Thus, from each state $(i,0), i\in\{0,\ldots m\}$, the \gls{mc} transitions to a state $(i,k),\,k\in\{0,\ldots,W_i-1\}$ with probability $p_t/W_i$ (dotted orange transitions in Fig.~\ref{fig:macro_mc}), while with probability $1-p_t$ the \gls{sta} accesses the channel. We identify such latter condition as being in a \emph{transmission state} (the \gls{mc} is in a state $(i,0)$ and attempts to transmit); when the other condition applies (transmission deferral) or the \gls{mc} is in a state $(i,k),\,k>0$, we say that the \gls{sta} is in a non-transmission state. As in~\cite{akhtar2018directional}, each transmission state is itself a \gls{mc}, which will be described in Sec.~\ref{sec:tx_state}; thus, in order not to generate confusion, we will refer to the \gls{mc} of Fig.~\ref{fig:macro_mc} as \emph{macro \gls{mc}}.

Let $p$ be the failure probability, which includes the collision probability (see the discussion in Sec.~\ref{sec:dir_comm}) and the error probability due to the wireless channel, as explained later.
Then, from state $(k,0)$ the \gls{mc} goes to a state $(0,i), i\in\{0,\ldots,W_0-1\}$ (successful transmission of a new packet; solid green transitions in Fig.~\ref{fig:macro_mc}) with probability $(1-p_t)(1-p)/W_0$, or to any state $(k+1,i), i\in\{0,\ldots,W_{k+1}-1\}$ with probability $(1-p_t)p/W_{k+1}$ (dashed red transitions in Fig.~\ref{fig:macro_mc}). If it reaches the maximum number of retransmission attempts ($k=m$), the \gls{mc} goes from state $(m,0)$ to a state $(0,i),\,i\in \{0,\ldots,W_0\}$  with probability $(1-p_t)/W_0$. 

It is then possible to compute the steady-state probabilities $\{b_{i,k}: i\!\in\!\{0,\ldots,m\}, k\!\in\!\{0,\ldots W_i-1\}\}$ of the macro \gls{mc}, using the same approach of~\cite{bianchi2000performance}. Assuming $p_t <1$, it is
\begin{equation}
 b_{i,k} = \frac{W_i-k}{W_i} p\,b_{0,0}\,,
\end{equation}
\setcounter{equation}{4}
where $b_{0,0}$ is given in~\eqref{eq:b00}.
Notice that $b_{0,0}$ does not depend on $p_t$, which, nonetheless, has an impact on the delay.

The probability of being in a transmission state is then
\begin{equation} \label{eq:tau}
 \tau = \sum_{i=0}^{m} b_{i,0} (1-p_t) = \frac{1-p^{m+1}}{1-p}(1-p_t) b_{0,0} \,.
\end{equation}

The time spent, on average, in a transmission or non transmission state is denoted as $\E{T_\ttx}$ and $\E{T_\nttx}$, respectively, which depend on the probabilities $\{b_{i,k}\}$ as explained next. It is then possible to define the probability $\pi_\ttx$ that, in an arbitrary time instant, the macro \gls{mc} is in a transmission state:
\begin{equation} \label{eq:pi_tx}
 \pi_\ttx = \frac{\tau \E{T_\ttx}}{\tau \E{T_\ttx} + (1-\tau) \E{T_\nttx}}\,.
\end{equation}
With probability $1-\pi_\ttx$, in an arbitrary time instant, the \gls{mc} will be in a non transmission state.
Note that~\eqref{eq:pi_tx} refers to the semi-Markov model, while~\eqref{eq:tau} refers to the corresponding embedded \gls{mc}.

To derive $\E{T_\ttx}$ and $\E{T_\nttx}$ it is first necessary to understand what happens in a transmission state.

\subsection{A transmission state} \label{sec:tx_state}

Whenever a \gls{sta} is in a transmission state $(i,0),\,i\in\{0,\ldots m\}$, it attempts to transmit with probability $1-p_t$, while with probability $p_t$ it picks a new backoff counter from the same window $W_i$ (see~\eqref{eq:p_t}) and enters a transmission state. 

%

To model such a behavior, each transmission state forms its own \gls{mc}, similarly to the model proposed in~\cite{akhtar2018directional}.
The \gls{mc} is made of $6$ states: the \emph{access} state $A$, the \emph{collision \gls{rts}} state $R_c$, the \emph{vulnerable \gls{rts}} state $R_v$, the \emph{ongoing transmission} state $O$, the \emph{failure} state $F$, and the \emph{success} state $S$. A \gls{sta} goes into state $A$ when it accesses the channel and goes from the macro \gls{mc} into the transmission state \gls{mc}. It transmits the \gls{rts} to the AP, and, based on the discussion in Sec.~\ref{sec:dir_comm}, two cases can occur.
\begin{itemize}
 \item As soon as the STA accesses the channel, it may immediately collide (case \ref{case:1} in Sec.~\ref{sec:dir_comm}). This happens with probability $p_{c,1}$; in this case, the \gls{sta} transitions to state $R_c$ where it transmits the \gls{rts} and then, with probability 1, goes to the failure state $F$.
 
 \item Otherwise, the transmission of the \gls{rts} is still vulnerable to interference. If it collides (case \ref{case:2} of Sec.~\ref{sec:dir_comm}), the \gls{mc} transitions to the failure state $F$; this happens with probability $p_{c,2}$. Otherwise, the \gls{sta} goes to state $O$, where it receives the \gls{cts} from the AP and then sends its data\footnote{No collisions can happen during the transmission of the CTS and we assume that there are no packet errors.}. In turn, the data transmission may fail because of channel errors (but not because of interference, as assumed in Sec.~\ref{sec:dir_comm}) and therefore, with probability $p_e$, the next state in the \gls{mc} is $F$. Otherwise, the transmission is successful and the next state in the \gls{mc} is $S$. Then, from either $F$ or $S$, the \gls{sta} exits the transmission state.
\end{itemize}

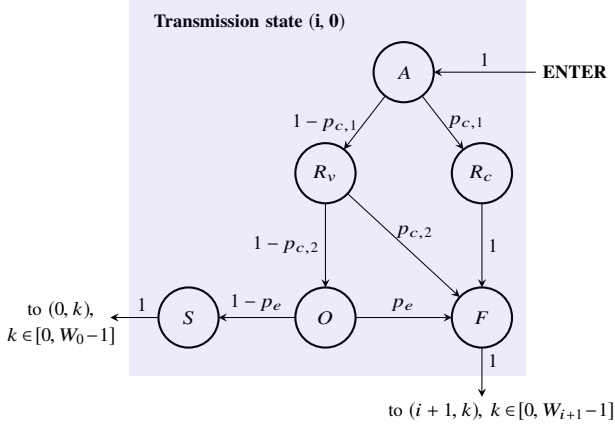
\begin{figure}[t]
	\centering
	\vspace{-1.5em}\definecolor{darkgreen}{rgb}{0,0.6,0.1}%
\definecolor{darkred}{rgb}{0.57,0.1,0}%
\definecolor{darkblue}{rgb}{0.2,0.1,0.7}%
\definecolor{darkorange}{rgb}{0.84,0.52}%
\definecolor{bluette}{rgb}{0.3,0.18,0.8}%

\begin{tikzpicture}[xscale=1.3,yscale=0.8,outer sep=0cm]
\scriptsize

\tikzset{>=stealth} 

\def \radius {.8cm}
\def \xspace {1cm}
\def \yspace {2.4cm}
\def \poshor {0.1}
\def \poscol {0.2}
\def \postimeout {0.2}

\node[draw, circle, minimum size=\radius,thick](A) at (0*\xspace,1.7*\yspace) {$A$};
\node[draw, circle, minimum size=\radius,thick](Rv) at (-0.8*\xspace,1*\yspace) {$R_v$};
\node[draw, circle, minimum size=\radius,thick](Rc) at (0.8*\xspace,1*\yspace) {$R_c$};
\node[draw, circle, minimum size=\radius,thick](O) at (-0.8*\xspace,0*\yspace) {$O$};
\node[draw, circle, minimum size=\radius,thick](F) at (0.8*\xspace,0*\yspace) {$F$};
\node[draw, circle, minimum size=\radius,thick](S) at (-2.2*\xspace,0*\yspace) {$S$};
\node[text width=3cm] (descr) at (-1.4*\xspace,2.05*\yspace) {\textbf{Transmission state} $\mathbf{(i,0)}$};

\draw[->] (A) to node[left] {$1-p_{c,1}$} (Rv);
\draw[->] (A) to node[right] {$p_{c,1}$} (Rc);
\draw[->] (Rc) to node[right] {$1$} (F);
\draw[->] (Rv) to node[above,xshift=.15cm] {$p_{c,2}$} (F);
\draw[->] (Rv) to node[left] {$1-p_{c,2}$} (O);
\draw[->] (O) to node[above] {$p_e$} (F);
\draw[->] (O) to node[above] {$1-p_e$} (S);
\draw[->] (S) to node[above,xshift=0.1cm] {$1$} (-3*\xspace,0*\yspace);
\draw[->] (1.35*\xspace,1.7*\yspace) to node[above] {$1$} (A);
\draw[->] (F) to node[right,yshift=.15cm] {$1$} (0.8*\xspace,-0.55*\yspace);

\draw [draw=none,fill=bluette, fill opacity=0.1] (-2*\xspace-\radius,-1.2*\radius) rectangle (1.25*\xspace,2.2*\yspace);

\node (success0) at (-3.5*\xspace,0.05*\yspace) {to $(0,k),$ };
\node (success2) at (-3.5*\xspace,-0.13*\yspace) {$k\!\in\![0,W_0\!-\!1]$};

\node (fail) at (1*\xspace,-0.65*\yspace) {to $(i+1,k),\,k\!\in\![0,W_{i+1}\!-\!1]$ };

\node (enter0) at (1.75*\xspace,1.7*\yspace) {\textbf{ENTER}};
%
\end{tikzpicture}\vspace*{-1.5em}
	\caption{Markov chain that models a transmission state. It is entered from i) state $(i,1)$ with probability $1-p_t$, ii) state $(i-1,0$) with probability $p(1-p_t)/W_0$ or iii) state $(i,0)$ itself with probability $p_t/W_0$.}
	\label{fig:micro_mc}
\end{figure}

The resulting \gls{mc} is represented in Fig.~\ref{fig:micro_mc}. 
Let $b_j$ be the steady-state probability that the \gls{mc} is in state $j\in \mathcal{J}_\ttx \triangleq \{A,R_c,R_v,O,F,S\}$. The transmission state itself forms a \gls{mc} where the outgoing transitions from states $S$ and $F$ re-enter the transmission state from state $A$. Thus, the steady-state probabilities are:  $b_A = 1/b_\ttx,\, b_{R_c} = (1-p_{c,1})/b_\ttx,\, b_{R_v} = p_{c,1}/b_\ttx,\, b_O = (1-p_{c,1})(1-p_{c,2})/b_\ttx, \, b_F = (1-(1-p_{c,1})(1-p_{c,2})(1-p_e))/b_\ttx, b_S = (1-p_e)(1-p_{c,1})(1-p_{c,2})/b_\ttx$, where $b_\ttx=3 + p_e(1-p_{c,1})(1-p_{c,2})$. 

Similarly to what done for the macro \gls{mc}, it is possible to define the probabilities $\pi_j$ that, in an arbitrary time instant, given that the \gls{mc} is in a transmission state, the \gls{mc} is in state $j\in \mathcal{J}_\ttx$:
\begin{equation} \label{eq:pi_j}
 \pi_j = \frac{T_j b_j}{\sum_{\ell\in\mathcal{J}_\ttx} T_\ell b_\ell} \quad j\in\mathcal{J}_\ttx\,,
\end{equation}
where $T_j$ is the time spent in state $j$.
Through $b_j$, the probabilities $\pi_j, j\in \mathcal{J}_\ttx$ depend on the collision and error probabilities $p_{c,1}$, $p_{c2,}$ and $p_e$. The collision probabilities in turn depend on how many and which other \glspl{sta} access the channel while the target \gls{sta} is in the transmission state, and thus on  $\pi_j$.

Considering the model in Sec.~\ref{sec:dir_comm} where \glspl{sta} can be grouped based on what they can hear of a communication between the \gls{ap} and another \gls{sta}, $p_{c,1}$ is given by the probability that none of these three cases occurs: i) any of the other STAs accesses the channel simultaneously, ii) at least a STA in group $n_{O,2}$ is transmitting an RTS to the AP, or iii) at least a STA in group $n_{O,4}$ is using the channel. We analyze the probabilities of these events.

Case i) occurs if at least another STA is accessing the channel, given that the target STA is accessing the channel. The probability of this to happen can be expressed as:
\begin{equation} \label{eq:q1}
\begin{aligned}
 q_1 & = \frac{1-(1-p_\acc)^n - n p_\acc(1-p_\acc)^{n-1}}{p_\acc}\,,
\end{aligned}
\end{equation}
where $n$ is the total number of \glspl{sta} in the network and $p_\acc = \pi_A \pi_\ttx$ is the probability that a STA is accessing the channel.

Case ii) happens if at least a \gls{sta} in group $n_{O,2}$ is either in state $R_v$ or in $R_c$, given that the target STA is accessing the channel.  Thanks to Bayes' rule, the probability of this to occur can be equivalently expressed as a function of the probability that the target STA accesses the channel given that at least a \gls{sta} in group $n_{O,2}$ is either in state $R_v$ or in $R_c$. 
This is not trivial to compute, because it requires an analytical expression for the relations between the coverage areas of multiple STAs. In fact, if a STA in group $n_{O,2}$ entered a transmission state, all STAs that can hear it refrain from transmitting, so that the number of STAs that compete for the channel is reduced and it is more likely that the target STA senses the channel as idle and attempts transmitting.
However, we do not consider such relations, which are extremely challenging to model, but only account for the fact that, if some STAs are in a transmission state, the ECDA operation is not frozen, so that the access probability is increased by a factor $1/(T_{\rm CBAP}/T_{\rm BI})$.
We thus express the probability of case ii) as
\begin{equation} \label{eq:q2}
 q_2 = \frac{1-(1-\pi_\ttx(\pi_{R_v}+\pi_{R_c}))^{n_{O,2}}}{T_{\rm CBAP}/T_{\rm BI}}\,.
\end{equation}
As the numerical evaluation of Sec.~\ref{sec:results} shows, this approximation affects the validity of the model only for highly dense scenarios, with more than $100$ STAs.

Finally, case iii) can be treated analogously to case ii) and thus happens with probability
\begin{equation} \label{eq:q3}
 q_3 =  \frac{1-(1-\pi_\ttx(\pi_{R_v}+\pi_{R_c}+\pi_O))^{n_{O,4}}}{T_{\rm CBAP}/T_{\rm BI}}\,.
\end{equation}

Then, the probability of colliding while accessing the channel is
\begin{equation} 
 p_{c,1} = 1- (1-q_1)(1-q_2)(1-q_3)\,. \label{eq:pc1} 
\end{equation}
since there is a collision if at least one of cases i), ii), iii) occurs.

When the target STA did not collide while attempting to transmit and is thus in state $R_v$, it collides if at least a \gls{sta} that cannot hear the uplink messages sent by the target \gls{sta} accesses the channel during the whole duration of $R_v$. Following the same reasoning as per~\eqref{eq:q2} and~\eqref{eq:q3}, this happens with probability
\begin{equation}
 p_{c,2} = \frac{1 - ((1-p_\acc)^{n_{I,2} + n_{I,4}})^{T_{R_v}/T_A}}{T_{\rm CBAP}/T_{\rm BI}} \label{eq:pc2} \,.
\end{equation}
%

Eqs.~\eqref{eq:pi_j}, \eqref{eq:pc1}, \eqref{eq:pc2} form a nonlinear system in the unknowns $\pi_j, p_{c,1}$ and $p_{c,2}$, which can be solved  using numerical techniques, as in Bianchi's original model.
The error probability $p_e$ instead depends on the \gls{snr} and the \gls{mcs} used. 

\section{Performance metrics} \label{sec:metric}

We evaluate the performance achievable in a \gls{cbap} in terms of throughput, delay and dropping rate. Before delving into their description, it is useful to derive the time spent in a transmission and non transmission state.

\subsection{Average time spent in a transmission state}

A STA that accesses a transmission state can follow $4$ different paths, depending on collision and errors.
The average time spent in a transmission state is thus the sum of the time associated to each of these paths, weighed for the probability of that path:
\begin{equation}
\begin{aligned}
 \E{T_\ttx} &= (T_A + T_{R_c} + T_F) p_{c,1} \\
 &+ (T_A + T_{R_v} + T_F) (1-p_{c,1})p_{c,2} \\
 &+ (T_A + T_{R_v} + T_O + T_F) (1-p_{c,1})(1-p_{c,2})p_e\\
 &+ (T_A + T_{R_v} + T_O + T_S) (1-p_{c,1})(1-p_{c,2})(1-p_e).
\end{aligned}
\end{equation}

This can be easily seen in Fig.~\ref{fig:micro_mc}.

The probabilities $\pi_j$ are defined in~\eqref{eq:pi_j} and the times $T_j$ are as follows: $T_A = \delta,\, T_{R_c} = RTS,\, T_{R_v} = RTS,\, T_O = CTS + \E{T_L} + ACK + 3 SIFS + 3 \delta, T_F = DIFS, T_S = DIFS$, 
%
where $\delta$ is the propagation delay, $RTS$ and $CTS$ represent the time needed to send an \gls{rts} and \gls{cts} message, respectively, $\E{T_L}$ is the average time needed to transmit a data packet, $ACK$ is the time to send an ACK, and $SIFS$ and $DIFS$ represent the \gls{sifs} and \gls{difs} durations, respectively~\cite{802.11ad}.

\subsection{Average time spent in a non-transmission state}

The time spent in a non-transmission state depends on what happens meanwhile: the \gls{cbap} may freeze, the target \gls{sta} may hear a transmission or sense the channel as idle.
The \gls{edca} mechanism assumes that the backoff counter is decremented only after the channel is sensed idle for a time slot of duration $\sigma$ (which is defined in the standard and depends on the PHY layer). 
Before that, the \gls{cbap} may freeze or be busy. 
We can interpret the freezing condition as a self-loop on a state $(i,k),\,k>0$ with probability
\begin{equation}
  p_f = 1 - \frac{T_{\rm CBAP}}{T_{\rm BI}}\,,
\end{equation}
so that on average $1/(1-p_f)$ iterations over $(i,k)$ are expected before a transition to $(i,k-1)$.

The target \gls{sta} senses the channel as idle when none of the \glspl{sta} in groups $n_{I,1}$ and $n_{I,3}$ is using the channel (i.e., is in any of the states $A, R_c, R_v, O$) and none of the \glspl{sta} is using the channel \emph{and} has already received a feedback from the AP (i.e., it is in state $O$):
\begin{equation} \label{eq:p_i}
 p_i = (1-\pi_\ttx(\pi_A+\pi_{R_c}+\pi_{R_v}+\pi_O))^{n_{I,1} + n_{I,3}}(1-\pi_\ttx\pi_O)^{n_{I,2}}\,,
\end{equation}


The channel is sensed as busy with probability $1-p_i$ for an average duration of $E_\ttx$.
Thus, the average time spent in a non-transmission state can be expressed as
\begin{equation} \label{eq:e_ntx}
 \E{T_\nttx} = \sigma + \frac{(1-p_i)E_\ttx}{1-p_f}\,.
\end{equation}

\subsection{Throughput}
The normalized system throughput $S$ is defined as the fraction of time that the channel is used to successfully transmit information.
The average payload size is $\E{L}$ and a transmission is successful with probability $\pi_\ttx(1-p)$. Thus the aggregated throughput is
\begin{equation}
 S = n \frac{\pi_\ttx(1-p)\E{L}}{\pi_\ttx\E{T_\ttx} + (1-\pi_\ttx)\E{T_\nttx}}
\end{equation}
where the denominator represents the average duration of a time slot and $n$ is the number of STAs in the network.

\subsection{Delay} 

The delay experienced by a (successfully transmitted) packet is the time elapsed from when it arrived at the \gls{mac} layer until it is received. 
Let $\E{D_i}$ denote the expected delay that a packet experiences when it is successfully transmitted at stage $i$, $\mathrm{TX(i)}$ the event of transmission at stage $i$, and $\mathrm{success}$ the event of a successful transmission. Then
\begin{equation}
 \E{D} =  \sum_{i=0}^{m} \prr(\mathrm{TX}(i) | \mathrm{success}) \E{D_i}
\end{equation}
where $\prr(\mathrm{TX}(i) | \mathrm{success})$ represents the probability that, given that a successful transmission happened, it was at stage $i$. The event $\mathrm{success}$ happens when the packet is not discarded after $m$ backoff stages, i.e., with probability $1-p^{m+1}$, since a packet is dropped if its transmission fails (with probability $p$) at stages $0,1,\dots,m$. Thus $\prr(\mathrm{TX}(i) | \mathrm{success}) = (1-p)p^i/(1-p^{m+1})$ since the packet was discarded at stages $0,1,\ldots,i-1$ and then successfully transmitted at stage $i$.
The term $\E{D_i}$ is the sum of the average backoff process delay in stages $0,1,\ldots,i$, the collision delay experienced in stages $0,1,\ldots,i-1$, and the time needed for the successful transmission at stage $i$. The first state $k$ in the $j^{\tth}$ backoff stage is uniformly distributed between $0$ and $W_j-1$; the counter is decremented until state $(j,0)$ ($k+1$ states are crossed) and then, with probability $p_t$ there is a transition back to a random state at stage $j$. Therefore, the delay term is:
\begin{equation}
\begin{aligned} \label{eq:Di}
 &\E{D_i} = i T_c + T_s + \E{T_\nttx} \sum_{j=0}^i \sum_{\ell=0}^{+\infty} p_t^{\ell} \sum_{k=0}^{W_j-1} \frac{k+1}{W_j}  \\
  &= i T_c + T_s + \frac{\E{T_\nttx}}{1-p_t}\sum_{j=0}^i \frac{W_j+1}{2}  \\
  &= i T_c + T_s + \frac{\E{T_\nttx}}{2(1-p_t)} (2^{\min(i,m')+1}\!-\!1\!+\!\max(i\!-\!m',0)2^{m'} )W_0 
\end{aligned}
\end{equation}
where $\E{T_\nttx}$ is the time spent in  a backoff state and $T_c$ and $T_s$ are the durations of a successful transmission and a collision, respectively:
\begin{align}
 T_s &= RTS + CTS + \E{T_L} + ACK + 3 SIFS + 4\delta\,, \label{eq:T_s}\\
 T_c &= RTS + DIFS + \delta \,.\label{eq:T_c}
\end{align}

\section{A model for directional communication} \label{sec:directional_model}

The model of Sec.~\ref{sec:dir_comm} assumes to know the number of \glspl{sta} that can overhear the uplink and downlink messages exchanged between the \gls{ap} and a target \gls{sta}, which is equivalent to characterizing the regions around the target \gls{sta} corresponding to groups $n_{I,1}$, $n_{I,2}$, $n_{I,3}$ and $n_{I,4}$.
This is not trivial to compute as the power received at a \gls{sta} depends on the gains of the transmitting and receiving antennas, as per~\eqref{eq:p_rx}, which vary according to the considered direction (angles $\theta_\ttx$ and $\varphi_\rrx$ in \eqref{eq:p_rx}). 
In the following, we describe the model we use for the beam shapes and then provide a mathematical approach to compute the areas corresponding to each group of \glspl{sta}.

\subsection{Beam shapes} \label{sec:beams}
The directivity of an antenna depends on the shape of a beam. There exists a multitude of models for antenna beams, such as the Gaussian beam shape, the sinc beam shape and the sampled beam shape, which, however, are very challenging to be used in mathematical models.
A simpler approach is given by the constant-gain  beam shape (sometimes called pizza-slice beam shape), 
where the space around the device is divided into $N_b$ beams with constant beamwidth $W_b=2\pi/N_b$; a beam has constant gain in the main lobe and there are no side lobes. From the expression of the directivity of an antenna~\cite{stutzman2012antenna}, the antenna gain for a beam centered at $\varphi$ is $g(\theta) =N_b$ if $\theta \in \left[\varphi - \frac{W_b}{2}, \varphi + \frac{W_b}{2}\right]$, and $0$ otherwise.


\begin{figure}[t]
  \centering
  \includegraphics[width=.45\columnwidth]{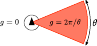}
  \caption{Pizza-slice beam of width $\theta$. The antenna gain is $2\pi/\theta$ within the beam, $0$ outside.}
    \label{fig:beam}
\end{figure}
We assume homogeneous \glspl{sta} with the same antenna gains; however it makes sense to consider the \gls{ap} to be more powerful than the \glspl{sta} and with narrower beams. Thus, we denote as $N_\ap$ and $N_\st$ the number of sectors for the \gls{ap} and a \gls{sta}, respectively. We assume that $N_\ap\ge 2$ and $N_\st \ge 2$. 

As explained in Sec.~\ref{sec:dir_comm}, since the \glspl{sta} only communicate with the \gls{ap}, they always have their transmitting and receiving antennas directed towards it.\footnote{Except of course during the beamforming training, which is however out of the scope of this paper.} The \gls{ap} instead listens in a \gls{qo} mode and switches to directional mode when engaged in a communication with a \gls{sta}. In this work, we assume that the \gls{qo} mode coincides with omnidirectionality and yields a unitary gain, and leave to future work the investigation of smaller widths. 

We also assume full transmitter/receiver reciprocity, meaning that a \gls{sta} uses the same sector to transmit to and receive from the \gls{ap}, and vice versa.
The antenna gains of the \gls{ap} and \gls{sta} in directional mode computed with the model in Sec.~\ref{sec:beams} are $g_\ap \equiv N_\ap$ and $g_\st \equiv N_\st$ in the main lobes, respectively, and zero outside (see Fig.~\ref{fig:beam}). 

As a final remark, we highlight that, as in most of the literature, we consider only 2D directivity, which highly simplifies the problem, although antennas clearly have a 3D radiation pattern.

\subsection{Coverage area and power regulations}
Since there are no side lobes, two \glspl{sta} can hear each other only if they are in each other's main lobe, respectively. 
Given this and considering the average, it is possible to derive a maximum transmission range by means of a threshold $\gamma_\tth$ on the \gls{snr} $\gamma = P_\rrx/N$, where $N$ is the noise power. Then, using~\eqref{eq:p_rx}, the distance $d$ between two devices should be 
\begin{equation} \label{eq:d}
 d \le \left(\frac{P_\ttx g_\ttx(\theta_\ttx,\varphi_\rrx)g_\rrx(\theta_\ttx,\varphi_\rrx)}{\gamma_\tth A N}\right)^{1/\eta}\,.
\end{equation}
The threshold $\gamma_\tth$ can be computed by imposing a maximum tolerable \gls{ber} and deriving the corresponding \gls{snr} (note that this depends on the \gls{mcs} used).
The antenna gains of the \gls{ap} and \gls{sta} are computed as described in Sec.~\ref{sec:beams}.

Interestingly, if the \gls{ap} and the \gls{sta} have different transmission powers, there is an \gls{snr} asymmetry between downlink and uplink when considering the same noise level at receiver and transmitter (see~\eqref{eq:p_rx} and~\eqref{eq:d}).
We assume the coverage radius $R$ to be bounded by the most stringent limit~\eqref{eq:d} between uplink ($P_\ttx$ and $g_\ttx$ are those of the \gls{sta}, $g_\rrx$ is that of the \gls{ap} which can be listening in either \gls{qo} or directional mode) and downlink communication ($P_\ttx$ and $g_\ttx$ are those of the \gls{ap}, $g_\rrx$ is that of the \gls{sta}).  
Then, we consider an area $\R=\pi R^2$ around the \gls{ap} and the \glspl{sta} uniformly distributed according to a Poisson Point Process of intensity $\lambda$.

\subsection{Stations that overhear uplink messages} \label{sec:ul}

We consider a Cartesian plane whose origin coincides with the center of area $\R$, so that the \gls{ap} is in $(0,0)$. 
Without loss of generality, we assume that the target \gls{sta} is in $(d_t,0), d_t \in [0,R]$.  
Considering the beam model of Sec.~\ref{sec:beams}, the interferer can overhear uplink communication from the target \gls{sta} to the \gls{ap} if it is in the main lobe of the target \gls{sta} and vice versa, otherwise the received power is $0$ as per~\eqref{eq:p_rx}.
Consider an interferer \gls{sta} at distance $d_i \in [0,R]$ from the \gls{ap}.
It can overhear the uplink communication if and only if the phase of its polar coordinates is in the range $[\varphi_{\rm lim}(d_i), 2\pi -\varphi_{\rm lim}(d_i)]$, where

\begin{equation} \label{eq:phi_lim}
\varphi_{\rm lim}(d_i) = \begin{cases}
                          & \pi -\dfrac{\theta_\st}{2} -\arcsin\left(\dfrac{d_i}{d_t}\sin\left(\dfrac{\theta_{\st}}{2}\right)\right)\;\text{if}  \;d_i \le d_t\\
                          & \pi -\dfrac{\theta_\st}{2} -\arcsin\left(\dfrac{d_t}{d_i}\sin\left(\dfrac{\theta_{\st}}{2}\right)\right) \;\text{if} \;d_i > d_t
                         \end{cases}
\end{equation}
The proof of this result is provided in Appendix~\ref{apx:phi_lim}.

Considering all possible distances $d_i$, we obtain the expected area of \glspl{sta} that can overhear uplink messages given the position of the target node $(d_t,0)$ as
\begin{equation} \label{eq:Rr}
\begin{aligned} 
 \R_R(d_t) & = \int_0^R \int_{\varphi_{\rm lim}(d_i)}^{2\pi-\varphi_{\rm lim}(d_i)} r \intd \theta r \, \intd r = \pi R^2 - 2\int_0^R \varphi_{\rm lim}(r) r \, \intd r \\
 &= \pi R^2 - 2\int_0^{d_t} \left(\pi -\frac{\theta_\st}{2} -\arcsin\left(\frac{r}{d_t}\sin\frac{\theta_{\st}}{2}\right) \right ) r \, \intd r \\
   &-2 \int_{d_t}^R \left(\pi -\frac{\theta_\st}{2} -\arcsin\left(\frac{d_t}{r}\sin\frac{\theta_{\st}}{2}\right) \right ) r \, \intd r\\
\end{aligned}
\end{equation}
which can be solved in closed form. 

The expected area of \glspl{sta} that can overhear uplink messages is obtained by averaging~\eqref{eq:Rr} over $d_t$:
\begin{equation} \label{eq:E_Rr}
\begin{aligned}
 \E{\R_R} & = \int_0^R \R_R(d_t) \frac{2 d_t}{R^2} \intd d_t
\end{aligned}
\end{equation}
which also can be solved in closed form and only depends on the beam width $\theta_\st$. 

Rigorously, the power received at \glspl{sta} that are too far from the target \gls{sta} is too small so that they should be excluded from $\R_R$; however, as usually done in the literature, we neglect this issue and leave it for future investigation. 

\begin{figure}[t]
  \centering
  \includegraphics[width=0.8\columnwidth]{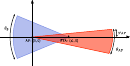}
  \caption{Location of the AP and the target STA in the Cartesian plane. The red beam is the AP beam in the direction of the STA, has width $\theta_\ap$ and goes from $\varphi_\ap-\theta_\ap$ to $\varphi_\ap$; the blue beam is the STA beam in the direction of the AP and has width $\theta_\st$.}
    \label{fig:regions}
\end{figure}

\subsection{Stations that overhear downlink messages} \label{sec:dl}

Without loss of generality we keep assuming that the target \gls{sta} is in $(d_t,0)$, and consider it to be in a random angular position within the \gls{ap} sector that covers it, which has width $\theta_\ap$. We thus denote as $\varphi_\ap\in[0,\theta_\ap]$ the angular phase of such sector, so that it spans the angles in the Cartesian plane in the range $[\varphi_\ap-\theta_\ap, \varphi_\ap]$, as shown in Fig.~\ref{fig:regions}. The covered area is 
\begin{equation} \label{eq:Rc}
 \R_C =  \int_0^R \int_{\varphi_\ap-\theta_\ap}^{\varphi_\ap} r \intd \theta r \, \intd r = \pi R^2 = \frac{\theta_\ap}{2} R^2\,.
\end{equation}
All \glspl{sta} in that sector can overhear downlink communication from the \gls{ap} to the target \gls{sta}, and, in particular, the \gls{cts}. Note that $\E{\R_C}\equiv \R_C$.

\subsection{Stations that overhear both uplink and downlink messages}
In this case, we have to consider the area that satisfies the requirements of both Secs.~\ref{sec:ul} and~\ref{sec:dl}. Thus
\begin{equation} \label{eq:Rrc}
\begin{aligned}
 \R_{R,C}(d_t,\varphi_\ap) & = \int_0^R \left(\int_{\varphi_{\rm lim}(r)}^{\varphi_\ap}\intd \theta + \int_{\varphi_\ap-\theta_\ap}^{2\pi-\varphi_{\rm lim}(r)}\intd \theta\right) r \, \intd r \,.
\end{aligned}
\end{equation}
Notice that $\varphi_{\rm lim}(\cdot)$ depends on the distance of the interferer from the \gls{ap}. It is possible to obtain a closed form expression for \eqref{eq:Rrc}, as explained in Appendix~\ref{apx:common_area}.

The corresponding expected area of \glspl{sta} that can overhear both uplink and downlink messages is obtained by averaging~\eqref{eq:Rrc} over $d_t\in [0,R]$ and $\varphi_\ap\in[0,\theta_\ap]$:
\begin{multline} \label{eq:E_Rrc}
 \E{\R_{R,C}} = \int_0^R \frac{2 d_t}{R^2} \Bigg(\int_{0}^{\theta_\ap}\frac{1}{\theta_\ap}\R_{R,C}(d_t,\varphi_\ap)\intd \phi_\ap \\
 +\int_{0}^{\theta_\ap}\frac{1}{\theta_\ap}\R_{R,C}(d_t,\varphi_\ap)\intd (\theta_\ap-\phi_\ap) \Bigg) \, \intd d_t\,.
\end{multline}
Appendix~\ref{apx:common_area} explains how to compute~\eqref{eq:E_Rrc}. Note that the integrals in~\eqref{eq:Rrc} yield zero for some positions of the target STA and the interferers. 

\subsection{Classification of the stations}

Given Eqs.~\eqref{eq:Rr}--\eqref{eq:E_Rrc}, it is possible to quantify the regions $\R_\ell, \ell\in\{1,2,3,4\}$ corresponding to the groups of nodes $n_{I,\ell}$ introduced in Sec.~\ref{sec:dir_comm}:
\begin{align}
 \R_1 &= \E{\R_{R}} - \E{\R_{R,C}}\\
 \R_2 &= \E{\R_{C}} - \E{\R_{R,C}}\\
 \R_3 &= \E{\R_{R,C}}\\
 \R_4 &= \R - \R_1 - \R_2 - \R_3 \,.
\end{align}

In this work, we assume that the \glspl{sta} are distributed according to a \gls{ppp}. Notice that, given the symmetry of the coverage areas, it is $n_{O,\ell} \equiv n_{I,\ell} \,\forall \ell$.

\section{Numerical evaluation} \label{sec:results}

\begin{table}[t]
 \centering
 \caption{Simulation parameters.}\vspace{-0.5em}
 \label{tab:params}
 \begin{tabular}{llr}
 \toprule
 \multicolumn{3}{l}{\textbf{BI structure}}\\
 BI duration & $BI$ & 100 ms\\
 BHI duration & $BHI$ & 2 ms\\
 \midrule
 \multicolumn{3}{l}{\textbf{EDCA parameters} \cite{802.11ad}}\\
 Minimum contention window size & $W_0$ & 16\\
 Maximum contention window size  & $2^{m'}W_0$ & $1024$\\
 Maximum $\#$ retransmission attempts & $m$ & {$6$}\\
 Slot duration & $\sigma$ & $5$ $\mu$s\\
 SIFS & $SIFS$ & $3$ $\mu$s\\
 DIFS & $DIFS$ & $13$ $\mu$s\\
 Propagation delay & $\delta$ & $100$ ns\\
 \midrule
 \multicolumn{3}{l}{\textbf{Packets size} \cite{802.11ad}}\\
 MAC header & $H_{MAC}$ & $320$ b\\
 PHY header & $H_{PHY}$ & $64$ b\\
 RTS size & $L_{RTS}$ & $20*8$ b\\
 CTS size & $L_{CTS}$ & $20*8$ b\\
 ACK size & $L_{ACK}$ & $14*8$ b\\
 Data size & $\E{L}$ & $7995\!*\!8 \text{ b } \!-\!H_{MAC}$ \\
 \midrule
 \multicolumn{3}{l}{\textbf{Noise}}\\
 Noise figure & $F_{dB}$ & $10$ dB\\
 Bandwidth & $W$ & $2.16$ GHz\\
 Path loss exponent & $\eta$ & $3$\\
 \bottomrule
 \end{tabular}
\end{table}

%

We validated the proposed model by comparing its performance in terms of throughput and delay with that of realistic Monte Carlo simulations for different system configurations.
In particular we investigate the accuracy of the model as a function of the fraction of DTI used for CBAP allocations, the number of such CBAP allocations, and the node density.

The system parameters are summarized in Table~\ref{tab:params}. 
The time required to send a message is computed as its size in bits (see Table~\ref{tab:params}) divided by the rate of the \gls{mcs} used; RTS, CTS and ACK messages are sent using the control modulation, which corresponds to a rate of $27.5$ Mb/s~\cite{802.11ad}, while we assume to use the Single Carrier PHY layer with $mcs=5$ for data transmission, which yields a data rate of $1251.25$ Mb/s~\cite{802.11ad}. 

We also set a maximum \gls{ber} of $10^{-6}$ and mapped such requirement onto a threshold $\gamma_\tth$ on the \gls{snr}\footnote{We built an SNR-BER map using the WLAN Toolbox\texttrademark of MATLAB software, which provides functions for modeling 802.11ad PHY.}. 
This allows to derive the area covered by the AP as per Sec.~\ref{sec:beams}, where the antenna gains are derived from the number of antenna sectors, and the noise power is $N\!=\!kT_0 F W$, where $k$ is Boltzmann constant and $T_0\!=\!290$ K; the noise figure $F$, the path loss exponent $\eta$ and the bandwidth $W$ are given in Table~\ref{tab:params}. The chosen configuration corresponds to a circular area of radius $R=23.5$ m.

\begin{figure}[t]
  \centering
  \setlength\fwidth{.9\columnwidth}
  \setlength\fheight{0.56\columnwidth}
  \vspace{-0.75em}\definecolor{mycolor4}{rgb}{0,0.9,0.65}%
\definecolor{mycolor3}{rgb}{0.85,0.38,0}%
\definecolor{mycolor2}{rgb}{0.92900,0.69400,0.12500}%
\definecolor{mycolor1}{rgb}{0.5,0,0.6}%
\definecolor{mycolor5}{rgb}{0.46600,0.67400,0.18800}%
\definecolor{mycolor6}{rgb}{0.30100,0.74500,0.93300}%
\definecolor{mycolor7}{rgb}{0.63500,0.07800,0.18400}%

\pgfplotsset{scaled x ticks=false}

\begin{tikzpicture}
\pgfplotsset{every tick label/.append style={font=\scriptsize}}
\tikzstyle{dotted} = [dash pattern=on \pgflinewidth off 0.5mm] 
\tikzstyle{dashed} = [dash pattern=on 7.5*0.8*0.8pt off 7.5*0.4*0.8pt]
\tikzstyle{dashed2} = [dash pattern=on 7.5*0.5*0.8pt off 7.5*0.25*0.8pt]
\tikzstyle{dashdotted} = [dash pattern=on 7.5*0.8*0.6pt off 7.5*0.8*0.3pt on \the\pgflinewidth off 7.5*0.8*0.3pt]
\tikzstyle{dotted2} = [dash pattern=on 7.5*0.8*0.2pt off 7.5*0.8*0.1pt]

\begin{axis}[%
width=0.951\fwidth,
height=\fheight,
at={(0\fwidth,0\fheight)},
scale only axis,
xmin=0.005,
xmax=0.055,
xtick={0.005,0.015,0.025,0.035,0.045,0.055},
xticklabels={0.005,0.015,0.025,0.035,0.045,0.055},
xlabel style={font=\color{white!15!black}},
xlabel={$\lambda$},
ymin=0,
ymax=0.6,
ytick={0,0.1,0.2,0.3,0.4,0.5,0.6},
ylabel style={font=\color{white!15!black}},
ylabel={Drop rate},
axis background/.style={fill=white},
legend style={font=\scriptsize,at={(0.03,0.97)}, anchor=north west, legend cell align=left, align=left, draw=white!15!black}
]
\addplot [color=mycolor1, line width=0.7pt]
  table[row sep=crcr]{%
0.005	0.0301553207844354\\
0.0075	0.0601407129725062\\
0.01	0.0915056744767177\\
0.0125	0.123062867245122\\
0.015	0.154292667492811\\
0.025	0.271378181194886\\
0.035	0.375041273110087\\
0.045	0.46765096204754\\
0.055	0.552015172475647\\
0.065	0.63105527973748\\
0.075	0.708094287610062\\
};
\addlegendentry{$\nu=0.25\:$ }

\addplot [color=mycolor2, line width=0.7pt]
  table[row sep=crcr]{%
0.005	0.0261482995769309\\
0.0075	0.0530121307425512\\
0.01	0.0811735277279558\\
0.0125	0.10961913801326\\
0.015	0.137890095262304\\
0.025	0.245061275558906\\
0.035	0.341966101977554\\
0.045	0.430671338040456\\
0.055	0.513648737635951\\
0.065	0.593652695238376\\
0.075	0.67408693143903\\
};
\addlegendentry{$\nu=0.5\:$ }

\addplot [color=mycolor3, line width=0.7pt]
  table[row sep=crcr]{%
0.005	0.0224473271735682\\
0.0075	0.0460591954421601\\
0.01	0.0709885276789287\\
0.0125	0.0962695939946596\\
0.015	0.121506761215734\\
0.025	0.218312198110871\\
0.035	0.30435578604865\\
0.045	0.404083211749867\\
0.055	0.473442825107997\\
0.065	0.554245925167846\\
0.075	0.638097945650065\\
};
\addlegendentry{$\nu=0.75\:$ }

\addplot [color=mycolor4, line width=0.7pt]
  table[row sep=crcr]{%
0.005	0.0189886131437167\\
0.0075	0.0395351204338471\\
0.01	0.0613257156134981\\
0.0125	0.0835074276792636\\
0.015	0.105748268550689\\
0.025	0.192109396644419\\
0.035	0.274081294368382\\
0.045	0.35359546281518\\
0.055	0.432762422674433\\
0.065	0.51416166004949\\
0.075	0.601374652243526\\
};
\addlegendentry{$\nu=1\:$ }

\end{axis}
\end{tikzpicture}%
\vspace{-1.5em}
  \caption{Drop rate vs STAs density for different values of $\nu$ when $N_{\rm CBAP}=N_{\rm SP}=3$.}
  \label{fig:res_lambda_drop}
\end{figure}
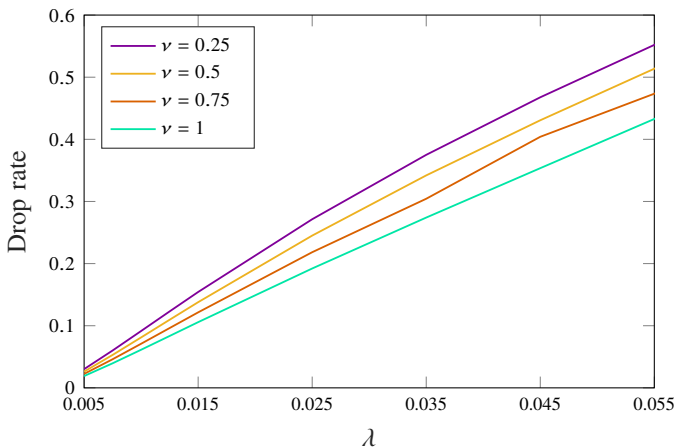

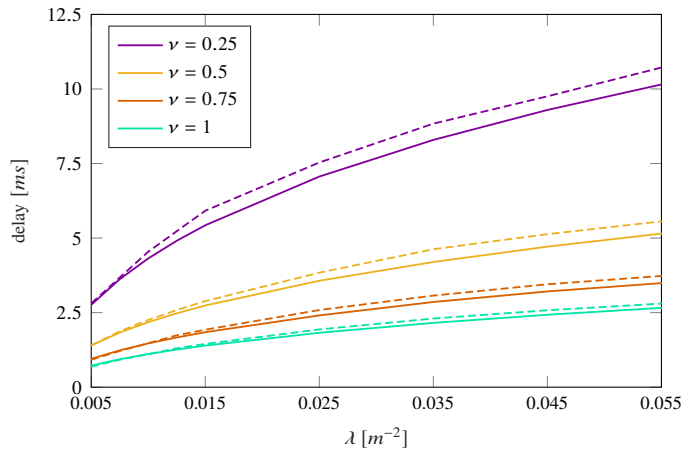
\begin{figure}[t]
  \centering
  \setlength\fwidth{.9\columnwidth}
  \setlength\fheight{0.56\columnwidth}
  \vspace{-0.75em}
%
%
\definecolor{mycolor4}{rgb}{0,0.9,0.65}%
\definecolor{mycolor3}{rgb}{0.85,0.38,0}%
\definecolor{mycolor2}{rgb}{0.92900,0.69400,0.12500}%
\definecolor{mycolor1}{rgb}{0.5,0,0.6}%
\definecolor{mycolor5}{rgb}{0.46600,0.67400,0.18800}%
\definecolor{mycolor6}{rgb}{0.30100,0.74500,0.93300}%
\definecolor{mycolor7}{rgb}{0.63500,0.07800,0.18400}%

\pgfplotsset{scaled x ticks=false}

\begin{tikzpicture}
\pgfplotsset{every tick label/.append style={font=\scriptsize}}
\tikzstyle{dotted} = [dash pattern=on \pgflinewidth off 0.5mm] 
\tikzstyle{dashed} = [dash pattern=on 7.5*0.8*0.8pt off 7.5*0.4*0.8pt]
\tikzstyle{dashed2} = [dash pattern=on 7.5*0.5*0.8pt off 7.5*0.25*0.8pt]
\tikzstyle{dashdotted} = [dash pattern=on 7.5*0.8*0.6pt off 7.5*0.8*0.3pt on \the\pgflinewidth off 7.5*0.8*0.3pt]
\tikzstyle{dotted2} = [dash pattern=on 7.5*0.8*0.2pt off 7.5*0.8*0.1pt]

\begin{axis}[%
width=0.951\fwidth,
height=\fheight,
at={(0\fwidth,0\fheight)},
scale only axis,
xmin=0.005,
xmax=0.055,
xtick={0.005,0.015,0.025,0.035,0.045,0.055},
xticklabels={0.005,0.015,0.025,0.035,0.045,0.055},
xlabel style={font=\scriptsize\color{white!15!black}},
xlabel={$\lambda\;[m^{-2}]$},
ymin=0,
ymax=12.5,
ytick={0,2.5,5,7.5,10,12.5},
ylabel style={font=\scriptsize\color{white!15!black}},
ylabel={delay $[ms]$},
axis background/.style={fill=white},
legend style={font=\scriptsize,at={(0.03,0.97)}, anchor=north west, legend cell align=left, align=left, draw=white!15!black}
]
\addplot [color=mycolor1, line width=0.7pt]
  table[row sep=crcr]{%
0.005	2.76259664901079\\
0.0075	3.62200896712574\\
0.01	4.31740408658111\\
0.0125	4.90950768951893\\
0.015	5.42902465143915\\
0.025	7.06241968996179\\
0.035	8.29166508480436\\
0.045	9.29550819002516\\
0.055	10.1509126286506\\
0.065	10.8971693173309\\
0.075	11.5549006264281\\
};
\addlegendentry{$\nu=0.25$}

\addplot [color=mycolor1, line width=0.7pt,dashed2, forget plot]
  table[row sep=crcr]{%
0.005	2.80900584416804\\
0.0075	3.67714034372703\\
0.01	4.54270860306345\\
0.0125	5.23100651336383\\
0.015	5.91656588589964\\
0.025	7.5341674203703\\
0.035	8.83709514348017\\
0.045	9.7530365836219\\
0.055	10.7175664632044\\
0.065	11.2866445875824\\
0.075	12.1044226959278\\
};

\addplot [color=mycolor2, line width=0.7pt]
  table[row sep=crcr]{%
0.005	1.39870316367506\\
0.0075	1.83217629500262\\
0.01	2.17994273181446\\
0.0125	2.47736643090452\\
0.015	2.73936329616066\\
0.025	3.56891715578604\\
0.035	4.19771861016029\\
0.045	4.7123585414369\\
0.055	5.15021789764852\\
0.065	5.53037340638599\\
0.075	5.86263067436944\\
};
\addlegendentry{$\nu=0.5$}

\addplot [color=mycolor2, line width=0.7pt,dashed2, forget plot]
  table[row sep=crcr]{%
0.005	1.39600989636615\\
0.0075	1.87456731817334\\
0.01	2.25492081668727\\
0.0125	2.58507755124352\\
0.015	2.88720134589492\\
0.025	3.84019429059531\\
0.035	4.62656407283528\\
0.045	5.13120564644297\\
0.055	5.56010325310244\\
0.065	5.96267598961769\\
0.075	6.24577230943177\\
};

\addplot [color=mycolor3, line width=0.7pt]
  table[row sep=crcr]{%
0.005	0.952107122547314\\
0.0075	1.23601989403107\\
0.01	1.46806593627364\\
0.0125	1.66747528408923\\
0.015	1.84388749426593\\
0.025	2.40676738179488\\
0.035	2.85673368641787\\
0.045	3.21043853460597\\
0.055	3.48970247496587\\
0.065	3.74874415534586\\
0.075	3.97312843613492\\
};
\addlegendentry{$\nu=0.75$}

\addplot [color=mycolor3, line width=0.7pt,dashed2, forget plot]
  table[row sep=crcr]{%
0.005	0.919925032090296\\
0.0075	1.21314756210972\\
0.01	1.47474838721224\\
0.0125	1.73913921174283\\
0.015	1.92769534406088\\
0.025	2.58776670688992\\
0.035	3.07140409490197\\
0.045	3.4508592546386\\
0.055	3.7295398448083\\
0.065	3.95424747012746\\
0.075	4.22414997492196\\
};

\addplot [color=mycolor4, line width=0.7pt]
  table[row sep=crcr]{%
0.005	0.725656922484705\\
0.0075	0.93738897622621\\
0.01	1.11141508668952\\
0.0125	1.26172806830944\\
0.015	1.39531318567743\\
0.025	1.82506884839375\\
0.035	2.15633716663307\\
0.045	2.42907336525166\\
0.055	2.66050000863804\\
0.065	2.85937766109661\\
0.075	3.03003164568341\\
};
\addlegendentry{$\nu=1$}

\addplot [color=mycolor4, line width=0.7pt,dashed2, forget plot]
  table[row sep=crcr]{%
0.005	0.688115171620091\\
0.0075	0.921124292889472\\
0.01	1.10486164262159\\
0.0125	1.30474604489052\\
0.015	1.45062166699035\\
0.025	1.93718653291651\\
0.035	2.30263898013001\\
0.045	2.57984982838207\\
0.055	2.80344278095405\\
0.065	3.00944125534208\\
0.075	3.14293489273898\\
};

\end{axis}
\end{tikzpicture}%
\vspace{-1.5em}
  \caption{Delay vs STAs density for different values of $\nu$. Analytical model (solid lines) vs. simulation (dashed lines) for $N_{\rm CBAP}=N_{\rm SP}=3$.}
  \label{fig:res_lambda_delay}
\end{figure}

Figs.~\ref{fig:res_lambda_S},~\ref{fig:res_lambda_drop} and~\ref{fig:res_lambda_delay} show the throughput, drop rate and delay, respectively, as functions of the STAs density $\lambda$ for different values of $\nu\triangleq T_{\rm CBAP}/T_{\rm DTI}$, which represents the fraction of DTI devoted to CBAP, as $T_{\rm DTI} = T_{\rm BI}-T_{\rm BHI}
= T_{\rm CBAP} + T_{\rm SP}$. Clearly, the throughput increases with $\nu$, since there is more time for EDCA operation. Interestingly, the simulations show that the STA density does not impact on the aggregated throughput: although the success rate of each STA considered separately decreases for larger values of $\lambda$, the number of STAs increases, yielding an almost constant value of $S$. The STAs density, however, does impact on the drop rate and on the delay.
The analytical model is more accurate for smaller values of $\nu$, while it tends to deviate from the simulated throughput as $\nu$ increases, because it overestimates the collision probability. Figs.~\ref{fig:res_lambda_S},~\ref{fig:res_lambda_drop} and~\ref{fig:res_lambda_delay} show the performance obtained with about $10$ STAs to up to about $100$ STAs in the network. As $\lambda$ increases, the model tends to underestimate the aggregated throughput. This happens because the collision probability is modeled by assuming the STAs to access the channel independently, while, as discussed in Sec.~\ref{sec:tx_state}, this is not true in reality. When the STAs density increases, the dependence among STAs becomes stronger but the model does not capture it and evaluates a poorer performance than that obtained in practice. 
The same considerations can be made for the delay (see Fig.~\ref{fig:res_lambda_delay}). The delay increases with the STA density because the higher collision rate leads to a larger number of retransmissions and decreases with $\nu$ since the EDCA operation is less likely to be frozen. The same effect can be seen for the drop rate in Fig.~\ref{fig:res_lambda_drop}.

\begin{figure}[t]
  \centering
  \setlength\fwidth{.9\columnwidth}
  \setlength\fheight{0.56\columnwidth}
  \vspace{-0.75em}
%
%
\definecolor{mycolor4}{rgb}{0,0.9,0.65}%
\definecolor{mycolor3}{rgb}{0.85,0.38,0}%
\definecolor{mycolor2}{rgb}{0.92900,0.69400,0.12500}%
\definecolor{mycolor1}{rgb}{0.5,0,0.6}%
\definecolor{mycolor5}{rgb}{0.46600,0.67400,0.18800}%
\definecolor{mycolor6}{rgb}{0.30100,0.74500,0.93300}%
\definecolor{mycolor7}{rgb}{0.63500,0.07800,0.18400}%

\pgfplotsset{scaled x ticks=false}

\begin{tikzpicture}
\pgfplotsset{every tick label/.append style={font=\scriptsize}}
\tikzstyle{dotted} = [dash pattern=on \pgflinewidth off 0.5mm] 
\tikzstyle{dashed} = [dash pattern=on 7.5*0.8*0.8pt off 7.5*0.4*0.8pt]
\tikzstyle{dashed2} = [dash pattern=on 7.5*0.5*0.8pt off 7.5*0.25*0.8pt]
\tikzstyle{dashdotted} = [dash pattern=on 7.5*0.8*0.6pt off 7.5*0.8*0.3pt on \the\pgflinewidth off 7.5*0.8*0.3pt]
\tikzstyle{dotted2} = [dash pattern=on 7.5*0.8*0.2pt off 7.5*0.8*0.1pt]

\begin{axis}[%
width=0.951\fwidth,
height=\fheight,
at={(0\fwidth,0\fheight)},
scale only axis,
xmin=0.005,
xmax=0.055,
xtick={0.005,0.015,0.025,0.035,0.045,0.055},
xticklabels={0.005,0.015,0.025,0.035,0.045,0.055},
xlabel style={font=\scriptsize\color{white!15!black}},
xlabel={$\lambda\;[m^{-2}]$},
ymin=0,
ymax=700,
ytick={0,100,200,300,400,500,600,700},
ylabel style={font=\scriptsize\color{white!15!black}},
ylabel={throughput $[Mb/s]$},
axis background/.style={fill=white},
legend style={font=\scriptsize,at={(0.05,0.975)}, anchor=north west, legend cell align=left, align=left, draw=white!15!black},
legend columns=4]
\addplot [color=mycolor1, line width=0.7pt]
  table[row sep=crcr]{%
0.005	127.576867787277\\
0.0075	130.817378014248\\
0.01	132.105131419704\\
0.0125	132.237813997295\\
0.015	131.615839117255\\
0.025	125.404579705614\\
0.035	116.237614070796\\
0.045	105.550460686572\\
0.055	93.8020204350021\\
0.065	80.9904875068685\\
0.075	66.7628246788625\\
};
\addlegendentry{$\nu=0.25\:$}

\addplot [color=mycolor1, line width=0.7pt,dashed2,forget plot]
  table[row sep=crcr]{%
0.005	132.576845588889\\
0.0075	134.4100665\\
0.01	135.5657689\\
0.0125	136.202893688889\\
0.015	136.64709205\\
0.025	136.5633259\\
0.035	135.018699761905\\
0.045	133.103297462687\\
0.055	129.676386928571\\
0.065	126.6300765\\
0.075	123.144578518519\\
};

\addplot [color=mycolor2, line width=0.7pt]
  table[row sep=crcr]{%
0.005	258.631154598711\\
0.0075	269.959071804773\\
0.01	274.806227383529\\
0.0125	276.760149622005\\
0.015	276.858986334765\\
0.025	268.070164001794\\
0.035	251.805939015259\\
0.045	231.434318230795\\
0.055	208.012345002181\\
0.065	181.517599386152\\
0.075	151.0989538206\\
};
\addlegendentry{$\nu=0.5\:$ }

\addplot [color=mycolor2, line width=0.7pt,dashed2,forget plot]
  table[row sep=crcr]{%
0.005	266.112392422222\\
0.0075	269.9266735\\
0.01	271.7144802\\
0.0125	272.902426866667\\
0.015	273.6451587\\
0.025	273.960771851852\\
0.035	270.7605166\\
0.045	266.498194105263\\
0.055	260.638061944444\\
0.065	253.646942777778\\
0.075	248.242794117647\\
};

\addplot [color=mycolor3, line width=0.7pt]
  table[row sep=crcr]{%
0.005	396.717858482965\\
0.0075	417.574844988172\\
0.01	428.683785083502\\
0.0125	434.49315546763\\
0.015	436.940980735737\\
0.025	430.000058223919\\
0.035	409.768286318453\\
0.045	371.9358117211\\
0.055	345.203989655298\\
0.065	303.924211542459\\
0.075	254.979682470152\\
};
\addlegendentry{$\nu=0.75\:$}

\addplot [color=mycolor3, line width=0.7pt,dashed2,forget plot]
  table[row sep=crcr]{%
0.005	398.977032233333\\
0.0075	404.931184740741\\
0.01	407.610585876543\\
0.0125	410.087452057613\\
0.015	411.196346320988\\
0.025	411.376654235845\\
0.035	406.697863588263\\
0.045	399.528848709315\\
0.055	391.951289391086\\
0.065	382.186519624102\\
0.075	369.741665608466\\
};

\addplot [color=mycolor4, line width=0.7pt]
  table[row sep=crcr]{%
0.005	537.484685777529\\
0.0075	563.788749427445\\
0.01	574.279732189573\\
0.0125	586.313215354823\\
0.015	593.009170126602\\
0.025	593.044519872179\\
0.035	570.422782207283\\
0.045	549.258366320167\\
0.055	507.404918612695\\
0.065	449.881324403944\\
0.075	379.599918447614\\
};
\addlegendentry{$\nu=1\:$ }

\addplot [color=mycolor4, line width=0.7pt,dashed2,forget plot]
  table[row sep=crcr]{%
0.005	532.048926311111\\
0.0075	540.080011466667\\
0.01	543.6100162\\
0.0125	546.775788\\
0.015	548.322482900763\\
0.025	548.2926474\\
0.035	542.2484384\\
0.045	534.138475\\
0.055	521.87982\\
0.065	507.19512057971\\
0.075	496.827431578947\\
};

\end{axis}
\end{tikzpicture}%
\vspace{-1.5em}
  \caption{Throughput vs STAs density for different values of $\nu$. Analytical model (solid lines) vs. simulation (dashed lines) for $N_{\rm CBAP}=N_{\rm SP}=3$.}
  \label{fig:res_lambda_S}
\end{figure}
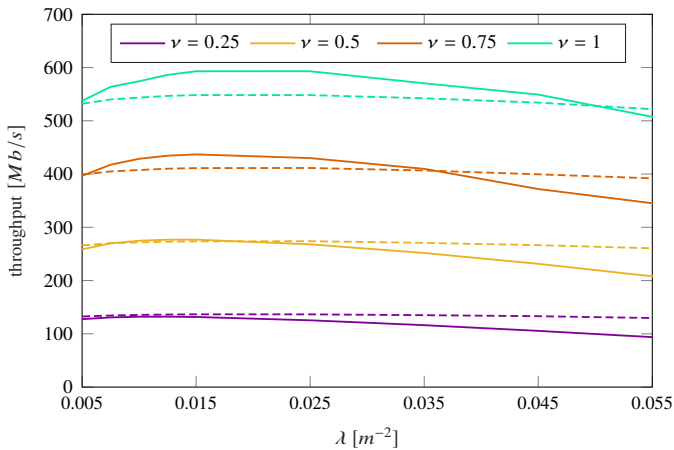

\begin{figure}[t]
  \centering
  \setlength\fwidth{.85\columnwidth}
  \setlength\fheight{0.5\columnwidth}
  \vspace{-0.75em}
%
%
\definecolor{mycolor4}{rgb}{0,0.9,0.65}%
\definecolor{mycolor3}{rgb}{0.85,0.38,0}%
\definecolor{mycolor2}{rgb}{0.92900,0.69400,0.12500}%
\definecolor{mycolor1}{rgb}{0.5,0,0.6}%
\begin{tikzpicture}
\pgfplotsset{every tick label/.append style={font=\scriptsize}}
\tikzstyle{dotted} = [dash pattern=on \pgflinewidth off 0.5mm] 
\tikzstyle{dashed} = [dash pattern=on 7.5*0.8*0.8pt off 7.5*0.4*0.8pt]
\tikzstyle{dashed2} = [dash pattern=on 7.5*0.5*0.8pt off 7.5*0.25*0.8pt]
\tikzstyle{dashdotted} = [dash pattern=on 7.5*0.8*0.6pt off 7.5*0.8*0.3pt on \the\pgflinewidth off 7.5*0.8*0.3pt]
\tikzstyle{dotted2} = [dash pattern=on 7.5*0.8*0.2pt off 7.5*0.8*0.1pt]

\begin{axis}[%
width=0.951\fwidth,
height=\fheight,
at={(0\fwidth,0\fheight)},
scale only axis,
xmin=0.1,
xmax=1,
xtick={0.1,0.2,0.3,0.4,0.5,0.6,0.7,0.8,0.9,1},
xlabel style={font=\scriptsize\color{white!15!black}},
xlabel={$\nu$},
ymin=0,
ymax=600,
ytick={0,100,200,300,400,500,600},
ylabel style={font=\scriptsize\color{white!15!black}},
ylabel={throughput $[Mb/s]$},
axis background/.style={fill=white},
legend style={font=\scriptsize,at={(0.03,0.97)}, anchor=north west, legend cell align=left, align=left, draw=white!15!black}
]
\addplot [color=mycolor1, line width=0.7pt]
  table[row sep=crcr]{%
0.1	42.1078921544299\\
0.2	87.1430582493141\\
0.3	135.354190800092\\
0.4	186.880893859179\\
0.5	241.865486107259\\
0.6	300.451904219703\\
0.7	362.78431355308\\
0.8	433.664440734378\\
0.9	499.254342833765\\
1	573.664460139364\\
};
\addlegendentry{$N_{\rm CBAP} = 1$}

\addplot [color=mycolor1, line width=0.7pt, dashed2, forget plot]
  table[row sep=crcr]{%
0.1	53.7659556875\\
0.2	107.736571043478\\
0.3	161.617104477612\\
0.4	215.185083466667\\
0.5	270.007070322581\\
0.6	323.686081973435\\
0.7	376.609816091954\\
0.8	431.049340727273\\
0.9	484.024626666667\\
1	537.958710769231\\
};

\addplot [color=mycolor2, line width=0.7pt]
  table[row sep=crcr]{%
0.1	43.0665613952751\\
0.2	88.0565777291195\\
0.3	136.217901194518\\
0.4	187.692433866381\\
0.5	242.62286678465\\
0.6	301.153251801347\\
0.7	363.427812666566\\
0.8	435.510029844532\\
0.9	499.776919294008\\
1	574.124113607648\\
};
\addlegendentry{$N_{\rm CBAP} = 10$}

\addplot [color=mycolor2, line width=0.7pt, dashed2, forget plot]
  table[row sep=crcr]{%
0.1	51.3254547096774\\
0.2	105.28164075\\
0.3	159.15488026534\\
0.4	212.355543\\
0.5	266.406014871795\\
0.6	321.231655028463\\
0.7	374.563519911012\\
0.8	427.536914133333\\
0.9	482.635614492754\\
1	536.522152258064\\
};

\end{axis}
\end{tikzpicture}%
\vspace{-1.5em}
  \caption{Throughput vs $\nu$ for two values of $N_{\rm CBAP}$. Analytical model (solid lines) vs. simulation (dashed lines) for $\lambda=0.04$ m$^{-2}$ and $N_{\rm SP}=10$.}
  \label{fig:res_frac_S}
\end{figure}
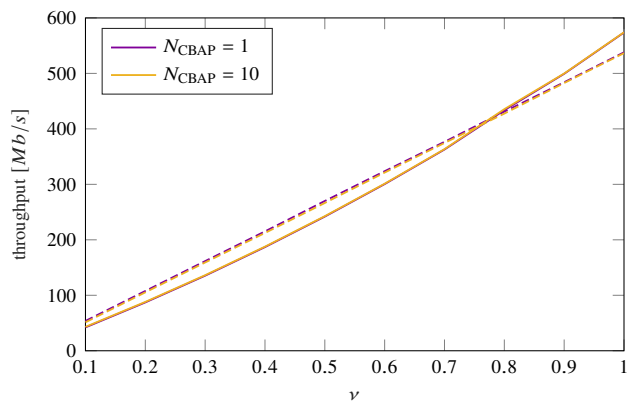

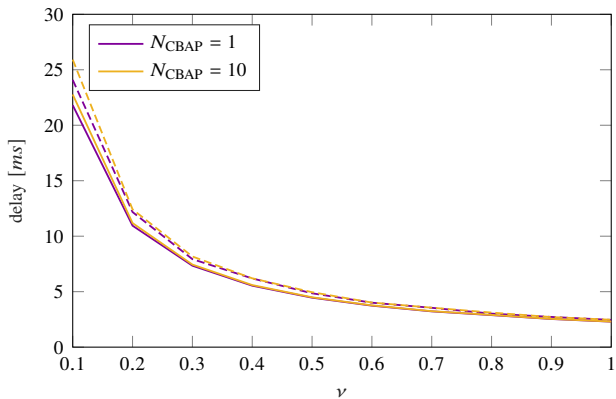
\begin{figure}[t]
  \centering
  \setlength\fwidth{.85\columnwidth}
  \setlength\fheight{0.5\columnwidth}
  \vspace{-0.75em}
%
%
\definecolor{mycolor4}{rgb}{0,0.9,0.65}%
\definecolor{mycolor3}{rgb}{0.85,0.38,0}%
\definecolor{mycolor2}{rgb}{0.92900,0.69400,0.12500}%
\definecolor{mycolor1}{rgb}{0.5,0,0.6}%
\begin{tikzpicture}
\pgfplotsset{every tick label/.append style={font=\scriptsize}}
\tikzstyle{dotted} = [dash pattern=on \pgflinewidth off 0.5mm] 
\tikzstyle{dashed} = [dash pattern=on 7.5*0.8*0.8pt off 7.5*0.4*0.8pt]
\tikzstyle{dashed2} = [dash pattern=on 7.5*0.5*0.8pt off 7.5*0.25*0.8pt]
\tikzstyle{dashdotted} = [dash pattern=on 7.5*0.8*0.6pt off 7.5*0.8*0.3pt on \the\pgflinewidth off 7.5*0.8*0.3pt]
\tikzstyle{dotted2} = [dash pattern=on 7.5*0.8*0.2pt off 7.5*0.8*0.1pt]

\begin{axis}[%
width=0.951\fwidth,
height=\fheight,
at={(0\fwidth,0\fheight)},
scale only axis,
xmin=0.1,
xmax=1,
xtick={0.1,0.2,0.3,0.4,0.5,0.6,0.7,0.8,0.9,1},
xlabel style={font=\scriptsize\color{white!15!black}},
xlabel={$\nu$},
ymin=0,
ymax=30,
ytick={0,5,10,15,20,25,30},
ylabel style={font=\scriptsize\color{white!15!black}},
ylabel={delay $[ms]$},
axis background/.style={fill=white},
legend style={font=\scriptsize,at={(0.03,0.97)}, anchor=north west, legend cell align=left, align=left, draw=white!15!black}
]
\addplot [color=mycolor1, line width=0.7pt]
  table[row sep=crcr]{%
0.1	21.805309003497\\
0.2	10.9490991651171\\
0.3	7.34138075915193\\
0.4	5.53938217123899\\
0.5	4.45864323085411\\
0.6	3.73824242934685\\
0.7	3.22363690747925\\
0.8	2.88661508424071\\
0.9	2.53724603114669\\
1	2.29685065202015\\
};
\addlegendentry{$N_{\rm CBAP} = 1$}

\addplot [color=mycolor1, line width=0.7pt, dashed2, forget plot]
  table[row sep=crcr]{%
0.1	24.085502866072\\
0.2	12.1766993723361\\
0.3	7.9328980164684\\
0.4	6.19302733964577\\
0.5	4.85322440442488\\
0.6	4.00505509901542\\
0.7	3.54507815223021\\
0.8	3.02038974448299\\
0.9	2.72253996084664\\
1	2.45529357274305\\
};

\addplot [color=mycolor2, line width=0.7pt]
  table[row sep=crcr]{%
0.1	22.7718120602492\\
0.2	11.1798496547456\\
0.3	7.44172635560108\\
0.4	5.5949222242937\\
0.5	4.49368962392877\\
0.6	3.76225773373862\\
0.7	3.24105211504284\\
0.8	2.90354664936709\\
0.9	2.54750944455452\\
1	2.3050541135197\\
};
\addlegendentry{$N_{\rm CBAP} = 10$}

\addplot [color=mycolor2, line width=0.7pt, dashed2, forget plot]
  table[row sep=crcr]{%
0.1	25.9150517675398\\
0.2	12.419147979267\\
0.3	8.16944346679207\\
0.4	6.20347493171602\\
0.5	4.95178913468728\\
0.6	4.01766053886724\\
0.7	3.55414637637768\\
0.8	3.1146091554216\\
0.9	2.71101801558877\\
1	2.46834463672161\\
};

\end{axis}
\end{tikzpicture}%
\vspace{-1.5em}
  \caption{Delay vs $\nu$ for two values of $N_{\rm CBAP}$. Analytical model (solid lines) vs. simulation (dashed lines) for $\lambda=0.04$ m$^{-2}$ and $N_{\rm SP}=10$.}
  \label{fig:res_frac_delay}
\end{figure}

In Figs.~\ref{fig:res_frac_S} and~\ref{fig:res_frac_delay} we evaluated the impact of $\nu$ and of the number of CBAP allocations $N_{\rm CBAP}$. We remember that $N_{\rm CBAP}$ impacts on $p_t$ (see~\eqref{eq:p_t}) since it determines the duration of each CBAP allocation for a given $T_{\rm CBAP}$. 
We fixed $N_{\rm SP}=10$ and varied $\nu$ for different values of $N_{\rm CBAP}$ with $\lambda=0.04$ m$^{-2}$. Interestingly, both the throughput and the delay do not depend on $N_{\rm CBAP}$ (the plots show only two values of $N_{\rm CBAP}$, but we obtained almost the same performance for each value of $N_{\rm CBAP}$ between $1$ and $N_{\rm SP}+1$).
The configuration of the CBAP allocation within a BI, thus, 
does not impact on the aggregated performance.
As already discussed, increasing $\nu$ improves the performance obtained during CBAP allocations.

Concluding, the proposed model provides results comparable to the simulations's one and, thus, can be used to easily analyze the impact of the system parameters on the performance obtainable in the CBAP allocations. Clearly, higher fractions of DTI allocated to \glspl{cbap} reduce the latency, improve the throughput and decrease the dropping rate. For a given $T_{\rm CBAP}$, however, the number of CBAP allocations within a BI, and thus the duration of each CBAP allocation, does not affect the aggregated performance. This information is useful because it allows to neglect the role played by $N_{\rm CBAP}$ when defining a configuration of the DTI that satisfies the application requirements, i.e., it is sufficient to choose only the total time $T_{\rm CBAP}$.

\section{Conclusion} \label{sec:conclusions}
In this work, we proposed an analytical model for the CBAP allocations in 802.11ad networks. We adapted the seminal work of Bianchi for legacy WiFi to account for the distinct features of the new amendment, including the interleaving of CBAP and SP allocations and the use of directional beams to communicate, which exacerbate the deafness and hidden node problems. 

Assessing the performance that can be obtained in CBAP allocations is the first step to design an efficient scheduler and determine the best DTI structure that accommodates the traffic requirements of multiple STAs. So, although the mathematical model slightly deviates from the real performance due to assumptions and simplifications, it is still able to capture the system behavior depending on the chosen configuration; thus, it can be extremely useful in the schedule planning phase and to gain insight on the impact of various parameters. 

In the next steps, we would like to characterize also the other types of allocations of 802.11ad, i.e., SPs and dynamic allocations, and merge such models to build an efficient scheduler.
We also would like to investigate what happens if the area around the AP is divided into regions that participate in different rounds of the CBAP operation.

Moreover, the model proposed in Sec.~\ref{sec:model} does not depend on the model used for the antenna beams, but for mathematical tractability we used the pizza-shaped beams. 
Naturally, in practical networks the antenna beams are not the ideal ``pencil-beams'', but are wider and irregular, with side lobes that are often neglected in the literature~\cite{assasa2018medium}; we may take into account some more realistic beam models in our future work.

Finally, we neglected the overhead needed for beamforming training and subsequent beam tracking, but it could be interesting to include it and analyze the impact of imperfect beam alignment on the communication performance, as well as the presence of obstacles in the communication paths. 

\appendices

\section{}  \label{apx:phi_lim}
\begin{figure}[t]
  \centering
  \includegraphics[width=.8\columnwidth]{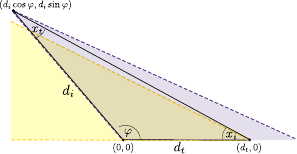}
  \caption{Target \gls{sta} and potential interferer in the Cartesian plane. The blue area represents the right half of the beam of the potential interference directed towards the AP in $(0,0)$; it has width $\theta_\st/2$ in correspondence of the potential interferer. Analogously, the yellow area is the upper half of the beam of the target STA. Notice that in this example the two \glspl{sta} cannot hear each other, since the antenna gain of the target STA in $(d_i\cos\varphi, d_i\sin\varphi)$ is zero; this means that $\varphi < \varphi_{\lim}(d_i)$.}
    \label{fig:phi_lim}
\end{figure}

Here we prove that \glspl{sta} in group $n_{I,1}$ have the phase of their polar coordinates in the range $[\varphi_{\rm lim}(d_i), 2\pi -\varphi_{\rm lim}(d_i)]$, with $\varphi_{\rm lim}$ given in~\eqref{eq:phi_lim}.

Consider a possible interferer at distance $d_i\in[0,R]$ from the AP and with angular phase $\varphi\in[0,\pi]$; this means that we are only focusing on the upper half of the circular area around the AP, being the scenario symmetric. Consider the triangle whose vertices are the AP $(0,0)$, the target STA $(d_t,0)$ and the interferer $(d_i \cos \varphi, d_i\sin \varphi)$, as in Fig.~\ref{fig:phi_lim}. The two edges that form the vertex coinciding with the AP have length $d_i$ and $d_t$, and the angle they form has width $\varphi$.
Denote the other two angles as $x_i$ and $x_t$, as in Fig.~\ref{fig:phi_lim}. 

We are interested in the angles $\varphi$ such that the target \gls{sta} is in the beam of the interferer, i.e., $x_i\le \theta_\st/2$, \emph{and} the interferer is in the beam of the target STA, i.e., $x_t\le \theta_\st/2$ (the beams are symmetric with respect to the AP and have width $\theta_\st$).
Moreover, the angles must satisfy the two following equations
\begin{align}
 & x_i + x_t + \varphi = \pi, \label{eq:sumtriangle} \\
 & \frac{d_i}{\sin x_i} = \frac{d_t}{\sin x_t}\,, \label{eq:sinelaw} 
\end{align}
where~\eqref{eq:sinelaw} comes from the law of sines. 
If $d_i \le d_t$, then $x_i\ge x_t$, and thus the limit condition is obtained for $x_i=\theta_\st/2$. Considering~\eqref{eq:sumtriangle} and~\eqref{eq:sinelaw}, we then obtain that we are interested in all angles $\varphi \ge \varphi_{\lim} = \pi - \theta_\st/2-\arcsin(d_i/d_t \sin \theta_\st/2)$. Similarly, when $d_i > d_t$, the limit condition is obtained for $x_t=\theta_\st/2$, yielding $\varphi_{\lim} = \pi - \theta_\st/2-\arcsin(d_t/d_i \sin \theta_\st/2)$. This proves Eq.~\eqref{eq:phi_lim}.
Taking into account also \glspl{sta} in the lower half of the area around the AP, we finally obtain that the other STAs can overhear the messages sent by the target \gls{sta} if and only if their phase is in $[\phi_{\lim}(d_i), 2\pi-\phi_{\lim}(d_i)]$, with $d_i$ being their distance from the AP.

\section{}  \label{apx:common_area}

Here we explain how to express~\eqref{eq:Rrc} and~\eqref{eq:E_Rrc} so as to compute them in closed form.
We focus only on the first of the two integrals in~\eqref{eq:Rrc}, since analogous considerations can be made for the second one, and refer to it as $I_1= \int_0^R \int_{\varphi_{\rm lim}(r)}^{\varphi_\ap}\intd \theta r \, \intd r.$
$I_1 $ is zero if $\varphi_{\rm lim}(r)\!>\!\varphi_\ap$ for the considered $r\in[0,R]$ and position $d_t$ of the target node (see~\eqref{eq:phi_lim}). In particular $I_1\!=\!0$ if
\begin{equation} \label{eq:I1=0}
 \arcsin\left(c_1  \sin\frac{\theta_\st}{2}\right) < \pi-\frac{\theta_\st}{2}-\varphi_\ap 
\end{equation}
with $c_1= r/d_t$ if $r\le d_t$ and $c_1= d_t/r$ otherwise (as per~\eqref{eq:phi_lim}).
We recall that $0\le\theta_\st\le\pi$ and $0\le\varphi_\ap\le\theta_\ap\le\pi$ by assumption. Therefore, denoting the sum $\pi-\theta_\st/2-\varphi_\ap$ as $c_2$, it is $-\pi/2\le c_2\le \pi$. 
Note also that $0\le c_1 \sin\theta_\st/2\le 1$, yielding $0\le \arcsin(c_1 \sin \theta_\st/2)\le\pi/2$. 
\begin{itemize}
 \item If $c_2 \in[\pi/2, \pi]$, then the conditions in~\eqref{eq:I1=0} are certainly true, yielding $I_1=0$. This happens if $\varphi_\ap < \pi/2-\theta_\st/2$, whatever the value of $d_t$.

 \item Otherwise $c_2 \in[-\pi/2, \pi/2]$. In this range the sine function is monotonically increasing, so that it is possible to apply it to both terms in~\eqref{eq:I1=0} without additional adjustment. This gives the condition $c_1 \sin\theta_\st/2< \sin\left(\pi-\theta_\st/2-\varphi_\ap\right)$, which can be expressed as $c_1 <  \sin\left(\pi-\frac{\theta_\st}{2}-\varphi_\ap\right)/\sin\frac{\theta_\st}{2} \triangleq c_3$. It follows that $I_1=0$ if $r<d_t c_3$ in the case $r\le d_t$ and if $r>d_t/c_3$ in the case $r> d_t$. This happens only if $c_3 \le 1$, i.e., $\phi_\ap < \pi-\theta_\st$.
\end{itemize}

Summing up, it is
\begin{equation} \label{eq:I1}
 \begin{aligned} &I_1 = \Bigg(\int\limits_{\max(d_t c_3,0)}^{d_t} \left(-c_2 + \arcsin\left(\frac{r}{d_t}\sin\frac{\theta_\st}{2}\right)\right) r \, \intd r  \\
 &+ \int_{d_t}^{\min(\frac{d_t}{c_3},R)}  \left(-c_2  + \arcsin\left(\frac{d_t}{r}\sin\frac{\theta_\st}{2}\right)\right) r \, \intd r \Bigg) \mathbbm{1}_{\varphi_\ap \ge\pi-\theta_\st}
 \end{aligned}
\end{equation}
with $\mathbbm{1}_X$ being the indicator function, equal to $1$ if condition $X$ is true, and to $0$ otherwise. 
The $\min$ and $\max$ operators in the integral limits ensure that the range $[0,R]$ is not exceeded.
Eq.~\eqref{eq:I1} can be solved in closed form: to this aim, it is necessary to evaluate the integration limits where there are the $\max$ and $\min$ operators.
The same procedure can be used to compute the second integral $I_2$ in~\eqref{eq:Rrc} using $\theta_\ap-\varphi_\ap$ rather than $\varphi_\ap$.
 
This expression of $\R_{R,C}(d_t,\varphi_\ap)$ can be used to compute its expectation as in~\eqref{eq:E_Rrc}. We can focus only on the first double integral (the one over $\phi_{\ap}$), since analogous considerations can be made on the second one (the one over $\theta_{\ap}-\phi_{\ap}$). Such first integral is made over $I_1$ and $I_2$. We consider only the integral over $I_1$ and denote it as $J_1$; the rest of the terms in~\eqref{eq:E_Rrc} can then be derived following the same approach.
It is necessary to characterize $d_t/c_3$ and $d_t c_3$ based on $d_t$ and $\varphi_\ap$, so as to remove the min and max operators in the terms in~\eqref{eq:I1}. 
It is $d_t c_3 > 0$ if $\varphi_\ap \le \pi-c$, and $d_t/c_3 < R$ if $d_t \le R c_3$.

Rigorously from~\eqref{eq:I1}, it is $J_1=0$ if $\phi_\ap < \pi-\theta_\st$. This can be checked beforehand as it only depends on system parameters.
It is then possible to evaluate $J_1$ in closed form and, 
repeating the same procedure for the other terms in~\eqref{eq:E_Rrc}, calculate $\E{\R_{R,C}}$.

\bibliographystyle{IEEEtran}
\bibliography{bib}

\end{document}